\documentclass[pre,preprint]{revtex4-1}
\usepackage{graphicx}
\usepackage[utf8]{inputenc}
\usepackage{color}

\usepackage{amssymb,amsmath}

\usepackage{hyperref}

\newcommand{\YB}[1]{#1}

\newcommand{\figwidth}{0.7\columnwidth}

\begin{document}

\title{Local elastic properties of polystyrene nanocomposites increase significantly due to non-affine deformations}

\date{\today}

\author{Yaroslav M. Beltukov}
\affiliation{Ioffe Institute, Politechnicheskaya Str. 26, 194021 St. Petersburg, Russia}
\author{Dmitry A. Conyuh}
\affiliation{Ioffe Institute, Politechnicheskaya Str. 26, 194021 St. Petersburg, Russia}
\author{Ilia A. Solov'yov}
\email{ilia.solovyov@uni-oldenburg.de}
\affiliation{Department of Physics, Carl von Ossietzky Universität Oldenburg, Carl-von-Ossietzky-Str. 9-11, 26129 Oldenburg, Germany}
\affiliation{Ioffe Institute, Politechnicheskaya Str. 26, 194021 St.  Petersburg, Russia}

\begin{abstract}
We investigate the local elastic properties of polystyrene doped with SiO$_2$ nanoparticles by analyzing the local density fluctuations. The density fluctuations were established from coarse-grained molecular dynamics simulations performed with the MARTINI force field. A significant increase in polystyrene stiffness was revealed within a characteristic range of 1.4~nm from the nanoparticle, while polystyrene density saturates to the bulk value at significantly shorter distances. The enhancement of the local elastic properties of the polymer was attributed to the effect of non-affine deformations at the length scale below 1~nm, which was further confirmed through the random matrix model with variable strength of disorder.
\end{abstract}
\maketitle

Polymer nanocomposites have attracted significant attention due to their unique properties and enormous potential as future materials~\cite{Mai-2006}. Experimental investigations have demonstrated that due to the nanoscale inclusions, polymer nanocomposites could possess tailored mechanical, electrical, and thermal properties, as compared to the pure polymers~\cite{Paul-2008, Vassiliou-2008, Knite-2002}. Among many characteristics, the elastic properties of pure polymers and associated nanocomposites have historically attracted considerable attention~\cite{Thostenson-2003, Rafiee-2009, Mesbah-2009}. It was established that polymer doping with nanoparticles even in small concentrations could lead to significant changes in the elasticity of the host material~\cite{Fu-2008, Ou-1998, Wang-2002, Wetzel-2003, Bershtein-2021}. For example, doping of polymethylmethacrylate with just 3 wt.~\% of SiO$_2$ nanoparticles may increase the storage modulus of the nanocomposite by 50~\%~\cite{Stojanovic-2009}.

It is well known that for a homogeneous elastic medium the classical elasticity theory defines the relation between local strain and local stress~\cite{Landau-1986}. In the case of doped polymers, the Eshelby theory can be used to determine the deformation of the nanoparticles and the surrounding medium caused by the macroscopic external stress~\cite{Eshelby-1957}. The resulting overall stiffness of nanocomposites can then be calculated using the Mori-Tanaka approach~\cite{Mori-1973, Benveniste-1987}. However, the Eshelby theory becomes inaccurate once the nanoparticles appear to be of the nanometer scale.

Elastic properties of nanocomposites have also been described by the so-called three-phase model~\cite{Odegard-2005}. The model assumes that the structure of a polymer is perturbed around the nanoparticle, which results in an effective interface layer around the nanoparticle with intermediate elastic properties. Due to the large total surface area of a nanoparticle, the interfacial layer has a strong influence on the overall stiffness of the nanocomposite. However, the properties and the thickness of the interfacial layer are generally unknown and at present the three-phase model was used phenomenologically~\cite{Bondioli-2005, Saber-2007, Qiao-2009, Wang-2011, Amraei-2019}.

Here we investigate local elastic properties of a polymer medium surrounding a nanoparticle and consider polystyrene doped with amorphous SiO$_2$ nanoparticles as an example. The results demonstrate a significant increase of polystyrene local elastic properties in a nanometer-large shell around the nanoparticles, where the internal structure of the polystyrene is only slightly perturbed. To elucidate the role of disorder in the studied system, we have further utilized the random matrix model with variable strength of disorder which permitted to attribute the observed effect to non-affine deformations in the polymer~\cite{Beltukov-2013}. The specific case study illuminates a general and versatile approach that can easily be adapted to study of arbitrary polymer nanocomposites.

An important property of a polymer nanocomposite is the inhomogeneity of its constituents at a microscopic length scale. The effect becomes especially significant for a polymer matrix since its constituent monomers have a much larger size than the individual atoms. Moreover, since polymers surrounding an embedded nanoparticle are often found in a glassy state, monomers' disorder of positions and orientations makes the polymer matrix even more inhomogeneous.

The presence of disorder leads to the so-called non-affine deformations under homogeneous external load, which could not be described by a combination of local stretches or shears. Non-affine deformations play a crucial role in macroscopic elasticity properties and were observed in many disordered solids; examples include metallic glasses~\cite{Jana-2019}, polymer hydrogels~\cite{Wen-2012}, supercooled liquids~\cite{Del-2008}, silica glass~\cite{Leonforte-2006}. In the case of Lennard-Jones glass it was shown that the classical elasticity theory description fails below a length scale of tens of molecular sizes~\cite{Leonforte-2005}.
Disorder in the polymer matrix is also expected to affect the influence of nanoparticles on the macroscopic elastic properties of the nanocomposite. It was particularly demonstrated recently that nanoparticles in a strongly disordered medium have a stronger impact on the macroscopic elastic properties than nanoparticles in an ordered matrix~\cite{Conyuh-2019}.

\begin{figure}[t]
\centering
\includegraphics[width=\figwidth]{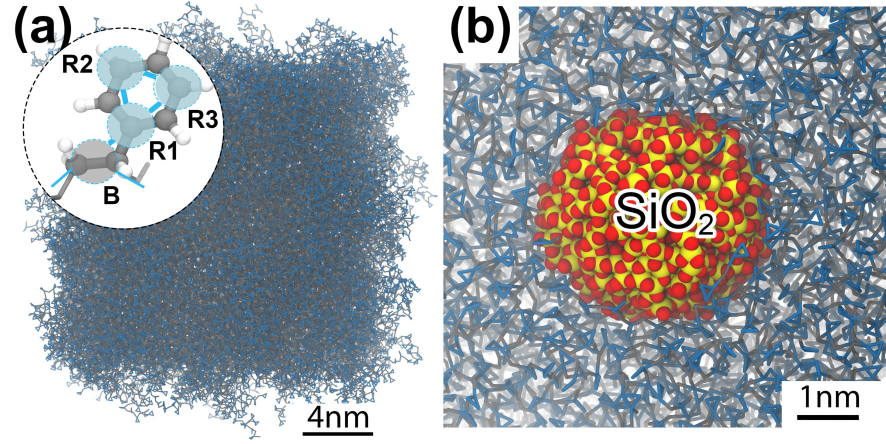}
\caption{(a) Characteristic configuration of pure polystyrene inside the simulation box obtained after MD equilibration~\cite{Beltukov2019}. Polystyrene is modeled in a coarse-grained representation, where each styrene monomer is represented through 4 beads denoted as particles R1, R2, R3, and B in the inset. (b) An illustrative configuration of the system featuring an amorphous SiO$_2$ nanoparticle embedded into the polystyrene matrix.}
\label{fig:PSMD}
\end{figure}

To study local elastic properties of polymer nanocomposites, two structures were considered in the present investigation: (i) a reference system of pure polystyrene, as illustrated in Fig.~\ref{fig:PSMD}(a) and polystyrene doped with an amorphous SiO$_2$ nanoparticle, see Fig.~\ref{fig:PSMD}(b). Pure polystyrene was modeled as a mixture of 216 polystyrene chains consisting of 120 monomers inside a 17.25~nm$\times$17.25~nm$\times$17.25~nm simulation box. The interaction of coarse-grained polystyrene beads was described by the MARTINI potential~\cite{Marrink2007,rossi2011coarse}, consistent with earlier studies~\cite{rossi2011coarse,Beltukov2019}. The polystyrene chains were coarse-grained to permit longer simulations and ensure equilibration of the system. Each styrene monomer was substituted with four coarse-grained beads, as illustrated in the inset to Fig.~\ref{fig:PSMD}(a). The backbone carbon atoms of polystyrene were modeled through one bead of type B, while three beads of type R represent the styrene side chain. After an extensive equilibration~\cite{Beltukov2019}, an additional 200~ns long molecular dynamics (MD) simulation was carried out at 300~K with the use of NAMD 2.13 software~\cite{PHIL2005}. The simulations assumed periodic boundary conditions within the NVT-statistical ensemble, where the temperature of the system was controlled through the Langevin thermostat with the damping coefficient of 1 ps$^{-1}$.

A single amorphous SiO$_2$ nanoparticle with a diameter of 3.6~nm was embedded into the equilibrated polystyrene matrix to model doped polystyrene. Accordingly, the simulation box was increased up to 17.285~nm$\times$17.285~nm$\times$17.285~nm to accommodate the nanoparticle and preserve the equilibrium density of the polystyrene matrix. An amorphous SiO$_2$ nanoparticle was constructed using the programs VMD~\cite{HUMP96}, and MBN Studio~\cite{Sushko2019} and embedded into the polystyrene matrix. Since periodic boundary conditions were employed in the simulation,
the studied system is essentially represented by an infinite medium with embedded nanoparticles, which corresponds to polystyrene doping with a mass fraction of 1.6\%. This value is consistent with the value typically used for nanocomposites~\cite{Stojanovic-2009}. The BKS-potential~\cite{Vollmayr1996} was used to describe the interatomic interactions inside the nanoparticle and its interaction with the polystyrene matrix. The doped polystyrene was simulated for 280~ns using the same parameters as for pure polystyrene. To ensure that the doped polystyrene systems reached equilibrium, and was suitable for the following analysis, the first simulated 80~ns were used to calculate and analyze the spatial distribution of polystyrene density in the simulation box. The approach was identical to the one used previously for pure polystyrene~\cite{Beltukov2019} and revealed an equilibrated polystyrene system with SiO$_2$ nanoinclusion.


To understand the macroscopic elastic properties of nanocomposites, it is essential to study local elastic moduli around the nanoparticles. One can, for example, analyze stress or strain fluctuations by means of the fluctuation-dissipation theorem~\cite{Sengupta-2000, Yoshimoto-2004,Tsamados-2009,Mizuno-2013, Sussman-2015}. The relation between elastic moduli and stress fluctuations contains the Born term, which relies on an involved analysis of the interaction potential between the constituents of the system~\cite{Amuasi-2015}. On the other hand, the strain fluctuations can readily be obtained directly from MD trajectories.

In this letter, we analyze the strain fluctuations of a polystyrene nanocomposite and determine its local elastic properties. We focus on studying the density fluctuations, which can be established without the knowledge of the precise equilibrium positions of atoms.

Let us define the relative density of the system at the position ${\bf r}$ and the time instance $t$ as
\begin{equation}
\xi({\bf r}, t) = \sum_i V_i\phi({\bf r}_i(t) - {\bf r}),
\label{eq:rho}
\end{equation}
where ${\bf r}_i(t)$ characterizes the position of the $i$th particle and $V_i$ is the volume of the Voronoi cell attributed to the $i$th particle and averaged over the observation time. The smoothing function $\phi({\bf r})$ reads as
\begin{equation}
\phi({\bf r}) = \frac{1}{(2\pi w^2)^{3/2}}\exp\left(-\frac{r^2}{2w^2}\right),   \label{eq:phi}
\end{equation}
where $w$ represents the spatial scale of the smoothing function. \YB{Thus, the parameter $w$ determines the spatial resolution of the relative density $\xi({\bf r}, t)$. This parameter plays an important role in the evaluation of the local elastic moduli by the analysis of the fluctuations of $\xi({\bf r}, t)$. For disordered media, such as polystyrene, bigger values of the spatial scale $w$ lead to a better precision of the elastic properties, but poorer spatial resolution. However, the resolution is important to analyze the elastic properties of nanoinclusions and the surrounding matrix. In the present paper, the dependence of local elastic moduli on the spatial scale $w$ will be analyzed.
}

The density $\xi({\bf r}, t)$ in Eq.~(\ref{eq:rho}) can be considered as a smoothed three-dimensional histogram with weights $V_i$. Since $V_i$ is the volume attributed to the $i$th particle, the density $\xi({\bf r}, t)$ is expected to be close to unity. Due to the thermal fluctuations, the density deviates from the average value, $\langle\xi({\bf r}, t) \rangle_t$, where the deviation of density is then defined as $\delta\xi({\bf r}, t) = \xi({\bf r}, t) - \langle\xi({\bf r}, t) \rangle_t$.

%
%

\begin{figure}[t!]
\centering
\includegraphics[width=\figwidth]{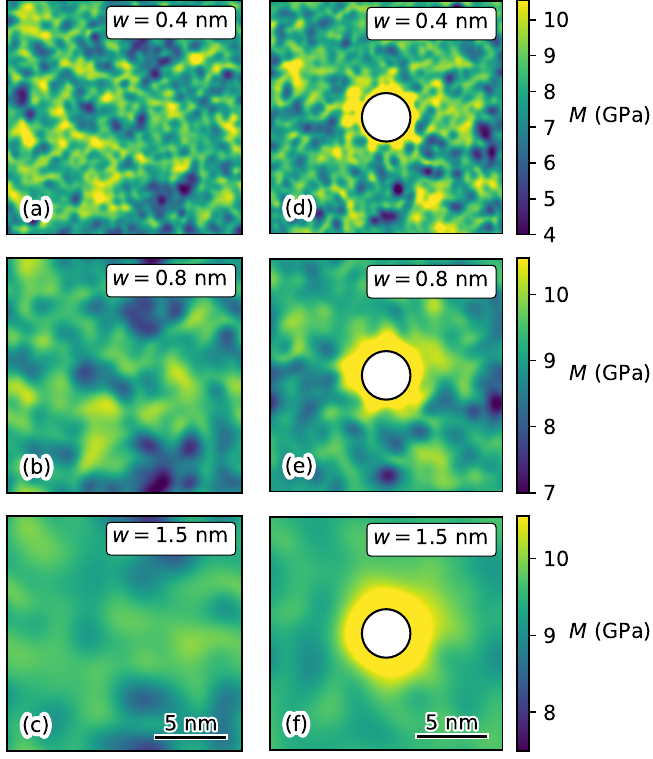}
\caption{Spatial distribution of the elastic modulus $M(\mathbf{r})$ for pure polystyrene (a--c) and doped polystyrene sample around a SiO$_2$ nanoinclusion of 3.6 nm diameter (d--f). The results are shown for the central cut-plane in the sample (for the doped sample it passes through the center of the nanoinclusion). The nanoinclusion is shown schematically with a circle. Results are shown for the three spatial scales $w=0.4$, $0.8$, and $1.5$~nm.}
\label{fig:map_PS}
\end{figure}

The thermal fluctuations of density allow determining the local elasticity modulus $M$ as:
\begin{equation}
M({\bf r}) = \left(\theta_3^3\bigl(e^{-4\pi^2 w^2/L^2}\bigr) - 1\right)\frac{k_B T \langle\xi({\bf r}, t) \rangle_t^2}{L^3 \langle\delta\xi^2({\bf r}, t) \rangle_t},   \label{eq:M}
\end{equation}
where $\theta_3(x)$ is the third Jacobi theta-function, $L$ defines the dimension of the simulation box, $k_B$ is the Boltzmann constant, and $T$ is the temperature of the system. The derivation of Eq.~(\ref{eq:M}) is given in the \hyperref[sm]{Supplemental Material} using thermal equilibrium analysis of a reference isotropic homogeneous elastic body. To the best of our knowledge, the relation between the local modulus $M({\bf r})$ and the local density fluctuations in Eq.~(\ref{eq:M}) was not derived before. For small spatial scales $w\ll L$ holds and Eq.~(\ref{eq:M}) can be further simplified as
\begin{equation}
M({\bf r}) = \frac{1}{8\pi^{3/2} }\frac{k_B T \langle\xi({\bf r}, t) \rangle_t^2}{w^3 \langle\delta\xi^2({\bf r}, t)\rangle_t}.   \label{eq:M2}
\end{equation}
The elasticity modulus $M$ is known as the P-wave modulus and determines the propagation velocity of longitudinal elastic waves as $v_l=\sqrt{M/\rho}$, where $\rho$ is the density~\cite{Mavko-1995}. It is related to other elastic moduli through the Poisson's ratio $\nu$~\cite{Mavko-2003}, which for glassy polymers is usually close to $1/3$~\cite{Gilmour-1979}.

The smooth relative density $\xi({\bf r}, t)$ was computed during the MD simulation of bulk polystyrene and polystyrene with an embedded nanoparticle. In the latter case, the sum in Eq.~(\ref{eq:rho}) includes atoms of the nanoparticle. Since the direct calculation of the convolution in Eq.~(\ref{eq:rho}) takes considerable calculation time, we use the smooth histogram method described in the \hyperref[sm]{Supplemental Material}.

\begin{figure}[t]
\centering
\includegraphics[width=\figwidth]{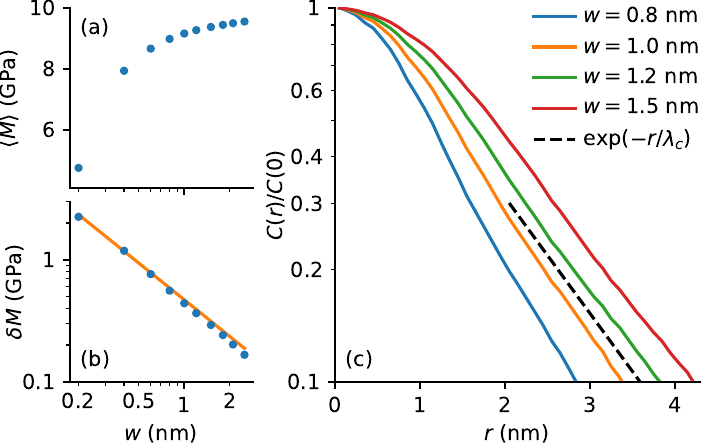}
\caption{(a) Mean $\langle M\rangle$ of the local elastic modulus for the doped polystyrene as a function of the spatial scale $w$. (b) Standard deviation $\delta M$ of the elastic modulus. The line shows the dependence $\delta M \propto 1/w$. (c) Correlation function of the local elastic modulus for the four spatial scales $w=0.8$, $1.0$, $1.2$, and $1.5$ nm. The dashed line shows the dependence $\exp(-r/\lambda_c)$, where $\lambda_c=1.4$ nm.}
\label{fig:corr}
\end{figure}

Figure~\ref{fig:map_PS}(a--c) shows the computed elastic modulus $M(\mathbf{r})$ for bulk polystyrene obtained for different spatial scales $w$. One notes strong spatial fluctuations of the elastic modulus in the case of $w=0.4$ nm, while the spatial fluctuations decrease for the larger $w$-values. The dependence of the mean elastic modulus $\langle M \rangle$ and the standard deviation $\delta M$ as a function of the spatial scale $w$ is presented in Fig.~\ref{fig:corr}(a--b). With increasing $w$, $\langle M \rangle$ tends to its macroscopic value and $\delta M$ decreases. For the investigated values of $w$ one observes $\delta M\propto 1/w$; in the case of uncorrelated noise one would expect that $\delta M$ will decrease with the increase of the smoothing length as $1/w^{3/2}$. It is thus important to study the correlation function $C({\bf r}) = \big\langle \big(M({\bf r}') - \langle M \rangle\big)\big(M({\bf r}' + {\bf r}) - \langle M \rangle\big)\big\rangle_{{\bf r}'}$ of the local elastic modulus to determine the relevant correlation length in the system. The obtained correlation function $C(r)$, averaged over the different spatial directions in the sample, is presented in Fig.~\ref{fig:corr} for several spatial scales $w$. One can see the exponential behavior of the correlation function $C(r) \propto \exp(-r/\lambda_c)$ with $\lambda_c \approx 1.4$ nm. Significant fluctuations of the local elastic modulus for small spatial scales and its correlated behavior make it necessary to study the nanoscale elastic properties in more detail. 



\begin{figure}[t]
\centering
\includegraphics[width=\figwidth]{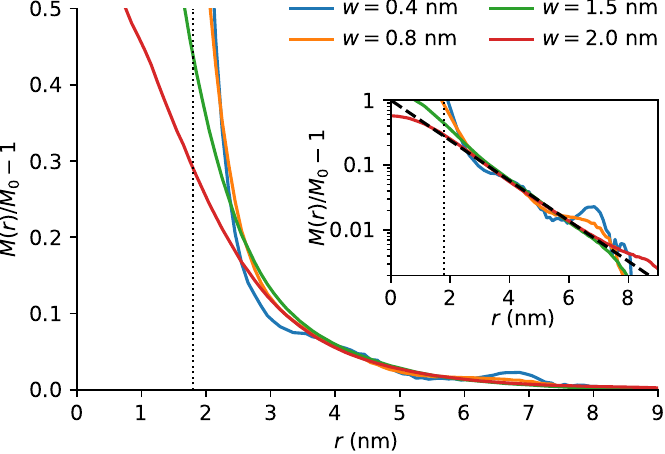}
\caption{Relative increase of the elastic modulus $M$ caused by the SiO$_2$ nanoinclusion computed as a function of distance from its center. Results are shown for the four considered spatial scales $w$. The vertical dotted line indicates the boundary of the nanoinclusion. The inset shows the same dependence in the logarithmic scale together with an asymptotic function $\exp(-r/\lambda)$, where $\lambda=1.4$ nm (dashed line).}
\label{fig:MR}
\end{figure}

Figure~\ref{fig:map_PS}(d--f) shows the elastic modulus $M(\mathbf{r})$ obtained for polystyrene with an embedded nanoparticle. Far from the nanoparticle, one observes fluctuations of the elastic modulus $M(\mathbf{r})$ similarly as for bulk polystyrene, see Fig.~\ref{fig:map_PS}(a--c). Inside the nanoparticle, the elastic modulus is significantly larger, $M=70$--$100$ GPa, corresponding to the typical values for SiO$_2$ \cite{Deschamps-2014, Mavko-1995} (see the \hyperref[sm]{Supplemental Material} for more details). It is, however, remarkable that the elastic modulus of polystyrene surrounding the nanoparticle is noticeably larger than the corresponding bulk value. The relative elastic modulus of polystyrene and the SiO$_2$ nanoparticle are analyzed in Fig.~\ref{fig:MR} as a function of distance from the nanoparticle center. A 10\% enhancement of the elastic modulus is observed for distances up to 1~nm from the nanoparticle surface. This result is not significantly influenced by the smoothing length scale $w$, as evidenced in the inset to Fig.~\ref{fig:MR}, which indicates that the observed enhancement of elastic modulus is not related to the smoothing of density fluctuations. Such elastic behavior is consistent with the description obtained from the three-phase model of the nanocomposite~\cite{Odegard-2005}, however, it is important to discuss the nature of the enhanced elastic properties around the nanoparticle. 

\begin{figure}[t]
\centering
\includegraphics[width=\figwidth]{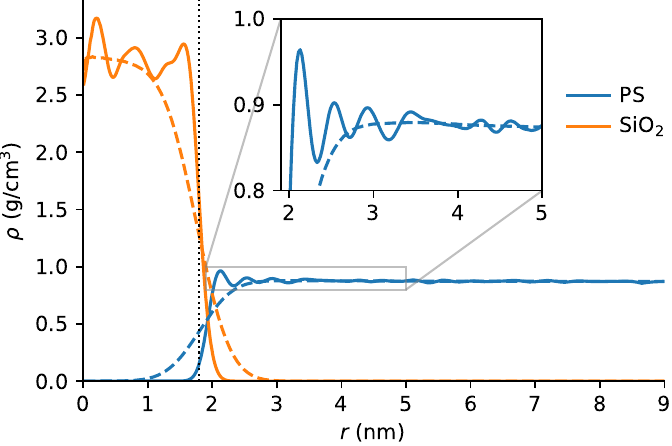}
\caption{Radial local density of the SiO$_2$ nanoparticle and the polystyrene (PS) matrix. The vertical dotted line marks the boundary of the nanoparticle. Solid and dashed lines show the density for $w=0.1$~nm and $w=0.4$~nm, respectively. Inset shows a zoom for PS density around the nanoparticle.}
\label{fig:density}
\end{figure}

The cause of the enhancement in the elastic properties in the vicinity of the nanoparticle can be analyzed through evaluating the density $\rho(\mathbf{r}) = \sum_i m_i \phi(\mathbf{r}-\mathbf{r}_i)$ separately for the nanoparticle and the polystyrene, as shown in Fig.~\ref{fig:density}. Here $m_i$ is the mass of the $i$th coarse-grained particle and $\phi$ is the smoothing function defined in Eq.~(\ref{eq:phi}). For the small smoothing value of $w=0.1$~nm, one notes clear oscillations of polystyrene density caused by coordination shells of the polystyrene monomers around the nanoparticle. The oscillations vanish for $w=0.4$~nm, and the polystyrene density approaches its bulk value with a deviation of less than 1\%. At the same time, Fig.~\ref{fig:MR} demonstrates a much larger deviation of $M(r)$, which shows that the enhancement of $M(r)$ near the nanoparticle is not directly related to the structural changes of the polymer in the vicinity of the nanoparticle. Other structural quantities also exhibit the same behavior near the nanoparticle and in the bulk polystyrene. Volumes of Voronoi cells around individual coarse-grain particles representing the polymer have homogeneous distribution except a narrow layer about 0.5 nm thick around the nanoparticle. Furthermore, the orientation of monomers is isotropic near the nanoparticle as well as in the bulk polystyrene. The detailed analysis of these quantities is presented in the \hyperref[sm]{Supplemental Material}.

In a strongly inhomogeneous medium, the local elastic properties are not completely determined by its local structure but depend on a large volume of the surrounding medium. Indeed, all atoms in the system tend to find new equilibrium positions for any applied macroscopic or microscopic stress. This deviation of equilibrium positions can be described by a continuous function ${\bf u}({\bf r})$; however, at the nanometer length scale, each atom has an additional non-affine displacement ${\bf u}_i^{\rm na}$. At the nanoparticle's surface, the non-affine displacements are suppressed because the nanoparticle is more homogeneous and stiffer than the surrounding polystyrene matrix, leading to different stiffness of the polystyrene matrix in the vicinity of the nanoparticle.

The inset in Fig.~\ref{fig:MR} shows that the additional stiffness of polystyrene has an exponential behavior $M(r)/M_0 - 1 \sim \exp(-r/\lambda)$ with $\lambda=1.4$~nm, where $M_0$ is the mean local elastic modulus far away from the inclusion ($r > 8.6$ nm) and the length scale $\lambda$ determines the characteristic length scale of non-affine displacements. The deviation from the exponential law in the inset in Fig.~\ref{fig:MR} is determined by small fluctuations of the local modulus $M(r)$, which remain after averaging of local modulus over a sphere with a given radius $r$. The obtained length scale $\lambda$ coincides with the correlation radius $\lambda_c$ of the obtained local modulus for the pure polystyrene (Fig.~\ref{fig:corr}). Note that the decay length $\lambda \approx 1.4$ nm for the shell of higher elastic modulus turns out to be close to the particle radius of 1.8~nm. The result suggests that a more systematic analysis of $\lambda$ on particle size is called for. Such an analysis is beyond the scope of this investigation as significant computations of polystyrene with nanoinclusions of different sizes would be necessary, and should be completed independently in a followup study.

The effect of disorder on local elastic properties around nanoparticles in an amorphous medium can be further quantified through employing the dimensionless random matrix model~\cite{Beltukov-2013}. The random matrix model describes many of the general vibrational and mechanical properties of amorphous solids~\cite{Beltukov-2013, Beltukov-2017, Conyuh-2021} and is based on the fundamental property that the dynamic matrix is positive definite for a system close to an equilibrium.

In the framework of the random matrix model, the disorder is controlled by the dimensionless parameter $\mu$, where $\mu \gg 1$ describes the regime with tiny fluctuations of the interaction between atoms; the other regime, $\mu \ll 1$ describes a strongly disordered amorphous material~\cite{Beltukov-2013}. To model a disordered medium around the nanoparticle one can consider a simple cubic lattice with the lattice constant $a_0$ and strong fluctuations of bond strength described by $\mu\ll1$. The nanoparticle is then described as a spherical region with $R<R_{\textsc{np}}$ and $\mu=\mu_{\textsc{np}}=1$ inside that region. The details of the random matrix model and its analysis are presented in the \hyperref[sm]{Supplemental Material}.

\begin{figure}[t]
\centering 
\includegraphics[width=\figwidth]{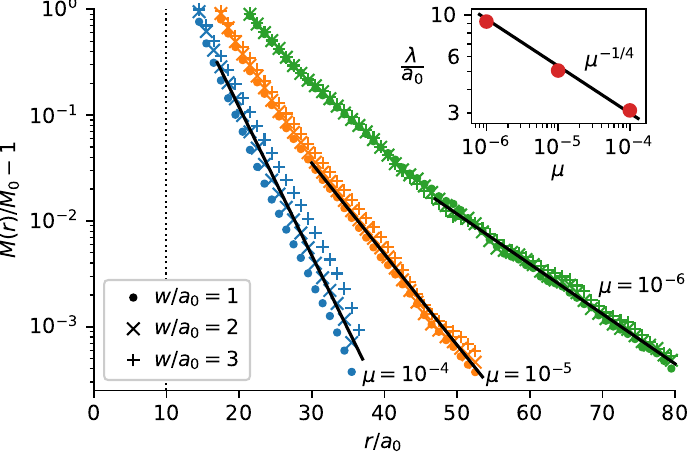}
\caption{Relative increase of the elastic modulus $M$ near the nanoinclusion computed within the random matrix model for different values of the spatial scale: $w=1a_0$ (dots), $w=2a_0$ (diagonal crosses), $w = 3a_0$ (vertical crosses). Different colors correspond to different values of the parameter $\mu=10^{-4}$, $10^{-5}$, $10^{-6}$. Dotted line shows the nanoinclusion size $R_{\textsc{np}}=10 a_0$. Solid lines show the exponential trend $\sim\exp(-r/\lambda)$ for the corresponding values of $\mu$. Inset shows the dependence of the fitted values of $\lambda$ on the parameter~$\mu$.}
\label{fig:dE}
\end{figure}

Following the random matrix model, the radial dependence of the calculated elastic modulus $M(r)$ is presented in Fig.~\ref{fig:dE} which reveals the exponential dependence of the elastic modulus around the nanoparticle on distance from its center. In the case of a stronger disorder (characterized by the smaller values of $\mu$), one notes a more prominent dependence that is not strongly dependent on the smoothing parameter $w$. For each value of the parameter $\mu$, one can determine the length scale $\lambda$ of the exponential decay of $M(r)/M_0 - 1 \sim \exp(-r/\lambda)$, such that $\lambda \sim \mu^{-\alpha}$ with $\alpha=0.24\pm 0.02$. The found dependency is close to $\lambda\sim\mu^{-1/4}$ (see inset to Fig.~\ref{fig:dE}), which is one of the known scaling regimes in the random matrix model~\cite{Beltukov-2013}. Since the random matrix model requires less computational resources than the MD simulation, it was used to verify whether the thickness $\lambda$ of the induced elastic shell is influenced by the radius of the nanoparticle. It was revealed that this is not the case but the thickness $\lambda$ was shown to be affected by the disorder of the surrounding matrix (see the \hyperref[sm]{Supplemental Material}).

The performed investigation concludes that an exponentially decreasing induced elastic shell is formed around a nanoparticle embedded inside a soft polystyrene matrix. Such a shell increases the effective volume of nanoinclusions inside polymeric materials, which leads to an increase in the effect on the macroscopic elastic properties of nanocomposites. 

The resulting elastic properties can approximately be modeled using the three-phase model with the interphase layer thickness $\lambda \approx 1.4$ nm. The observed enhancement of polystyrene's  elastic properties could not be explained by the deviation of its density and other structural properties. The result is consistent with a recent study for boehmite nanolayer in epoxy~\cite{Fankhanel-2019}. We conclude that the increase of elastic properties of the polystyrene matrix was caused by its non-affine deformations, which are strongly inhomogeneous at the nanometer length scale. This conclusion was supported by the analysis within the random matrix model with variable strength of disorder. We expect the maximum effect on macroscopic elastic properties if a typical distance between nanoparticles is of the order of the length scale $\lambda$ and the induced elastic shells do not overlap significantly.

The presented results show that detailed elasticity maps at different length scales can be computed by the analysis of the density fluctuations, which can be performed on the fly during an MD simulation. The same analysis can be applied to study local elastic properties in various soft inhomogeneous mediums and liquids as well.

\begin{acknowledgments}
The financial support from the Russian Science Foundation under the grant \# 17-72-20201 is gratefully acknowledged.
\end{acknowledgments}

\nocite{Gronbech-2020, Deschamps-2014}

\bibliographystyle{apsrev4-1}
\bibliography{bib/journals_short,bib/IS_group,bib/localmoduli,bib/ilia}

\begin{thebibliography}{51}%
\makeatletter
\providecommand \@ifxundefined [1]{%
 \@ifx{#1\undefined}
}%
\providecommand \@ifnum [1]{%
 \ifnum #1\expandafter \@firstoftwo
 \else \expandafter \@secondoftwo
 \fi
}%
\providecommand \@ifx [1]{%
 \ifx #1\expandafter \@firstoftwo
 \else \expandafter \@secondoftwo
 \fi
}%
\providecommand \natexlab [1]{#1}%
\providecommand \enquote  [1]{``#1''}%
\providecommand \bibnamefont  [1]{#1}%
\providecommand \bibfnamefont [1]{#1}%
\providecommand \citenamefont [1]{#1}%
\providecommand \href@noop [0]{\@secondoftwo}%
\providecommand \href [0]{\begingroup \@sanitize@url \@href}%
\providecommand \@href[1]{\@@startlink{#1}\@@href}%
\providecommand \@@href[1]{\endgroup#1\@@endlink}%
\providecommand \@sanitize@url [0]{\catcode `\\12\catcode `\$12\catcode
  `\&12\catcode `\#12\catcode `\^12\catcode `\_12\catcode `\%12\relax}%
\providecommand \@@startlink[1]{}%
\providecommand \@@endlink[0]{}%
\providecommand \url  [0]{\begingroup\@sanitize@url \@url }%
\providecommand \@url [1]{\endgroup\@href {#1}{\urlprefix }}%
\providecommand \urlprefix  [0]{URL }%
\providecommand \Eprint [0]{\href }%
\providecommand \doibase [0]{http://dx.doi.org/}%
\providecommand \selectlanguage [0]{\@gobble}%
\providecommand \bibinfo  [0]{\@secondoftwo}%
\providecommand \bibfield  [0]{\@secondoftwo}%
\providecommand \translation [1]{[#1]}%
\providecommand \BibitemOpen [0]{}%
\providecommand \bibitemStop [0]{}%
\providecommand \bibitemNoStop [0]{.\EOS\space}%
\providecommand \EOS [0]{\spacefactor3000\relax}%
\providecommand \BibitemShut  [1]{\csname bibitem#1\endcsname}%
\let\auto@bib@innerbib\@empty
\bibitem [{\citenamefont {Mai}\ and\ \citenamefont {Yu}(2006)}]{Mai-2006}%
  \BibitemOpen
  \bibfield  {author} {\bibinfo {author} {\bibfnamefont {Y.}~\bibnamefont
  {Mai}}\ and\ \bibinfo {author} {\bibfnamefont {Z.}~\bibnamefont {Yu}},\
  }\href@noop {} {\emph {\bibinfo {title} {Polymer Nanocomposites}}},\ Woodhead
  Publishing Series in Composites Science and Engineering\ (\bibinfo
  {publisher} {Elsevier Science},\ \bibinfo {year} {2006})\BibitemShut
  {NoStop}%
\bibitem [{\citenamefont {Paul}\ and\ \citenamefont
  {Robeson}(2008)}]{Paul-2008}%
  \BibitemOpen
  \bibfield  {author} {\bibinfo {author} {\bibfnamefont {D.~R.}\ \bibnamefont
  {Paul}}\ and\ \bibinfo {author} {\bibfnamefont {L.~M.}\ \bibnamefont
  {Robeson}},\ }\href@noop {} {\bibfield  {journal} {\bibinfo  {journal}
  {Polymer}\ }\textbf {\bibinfo {volume} {49}},\ \bibinfo {pages} {3187}
  (\bibinfo {year} {2008})}\BibitemShut {NoStop}%
\bibitem [{\citenamefont {Vassiliou}\ \emph {et~al.}(2008)\citenamefont
  {Vassiliou}, \citenamefont {Bikiaris}, \citenamefont {Chrissafis},
  \citenamefont {Paraskevopoulos}, \citenamefont {Stavrev},\ and\ \citenamefont
  {Docoslis}}]{Vassiliou-2008}%
  \BibitemOpen
  \bibfield  {author} {\bibinfo {author} {\bibfnamefont {A.}~\bibnamefont
  {Vassiliou}}, \bibinfo {author} {\bibfnamefont {D.}~\bibnamefont {Bikiaris}},
  \bibinfo {author} {\bibfnamefont {K.}~\bibnamefont {Chrissafis}}, \bibinfo
  {author} {\bibfnamefont {K.}~\bibnamefont {Paraskevopoulos}}, \bibinfo
  {author} {\bibfnamefont {S.}~\bibnamefont {Stavrev}}, \ and\ \bibinfo
  {author} {\bibfnamefont {A.}~\bibnamefont {Docoslis}},\ }\href@noop {}
  {\bibfield  {journal} {\bibinfo  {journal} {Composites Science and
  Technology}\ }\textbf {\bibinfo {volume} {68}},\ \bibinfo {pages} {933}
  (\bibinfo {year} {2008})}\BibitemShut {NoStop}%
\bibitem [{\citenamefont {Knite}\ \emph {et~al.}(2002)\citenamefont {Knite},
  \citenamefont {Teteris}, \citenamefont {Polyakov},\ and\ \citenamefont
  {Erts}}]{Knite-2002}%
  \BibitemOpen
  \bibfield  {author} {\bibinfo {author} {\bibfnamefont {M.}~\bibnamefont
  {Knite}}, \bibinfo {author} {\bibfnamefont {V.}~\bibnamefont {Teteris}},
  \bibinfo {author} {\bibfnamefont {B.}~\bibnamefont {Polyakov}}, \ and\
  \bibinfo {author} {\bibfnamefont {D.}~\bibnamefont {Erts}},\ }\href@noop {}
  {\bibfield  {journal} {\bibinfo  {journal} {Materials Science and
  Engineering: C}\ }\textbf {\bibinfo {volume} {19}},\ \bibinfo {pages} {15}
  (\bibinfo {year} {2002})}\BibitemShut {NoStop}%
\bibitem [{\citenamefont {Thostenson}\ and\ \citenamefont
  {Chou}(2003)}]{Thostenson-2003}%
  \BibitemOpen
  \bibfield  {author} {\bibinfo {author} {\bibfnamefont {E.~T.}\ \bibnamefont
  {Thostenson}}\ and\ \bibinfo {author} {\bibfnamefont {T.-W.}\ \bibnamefont
  {Chou}},\ }\href@noop {} {\bibfield  {journal} {\bibinfo  {journal} {Journal
  of Physics D: Applied Physics}\ }\textbf {\bibinfo {volume} {36}},\ \bibinfo
  {pages} {573} (\bibinfo {year} {2003})}\BibitemShut {NoStop}%
\bibitem [{\citenamefont {Rafiee}\ \emph {et~al.}(2009)\citenamefont {Rafiee},
  \citenamefont {Rafiee}, \citenamefont {Wang}, \citenamefont {Song},
  \citenamefont {Yu},\ and\ \citenamefont {Koratkar}}]{Rafiee-2009}%
  \BibitemOpen
  \bibfield  {author} {\bibinfo {author} {\bibfnamefont {M.~A.}\ \bibnamefont
  {Rafiee}}, \bibinfo {author} {\bibfnamefont {J.}~\bibnamefont {Rafiee}},
  \bibinfo {author} {\bibfnamefont {Z.}~\bibnamefont {Wang}}, \bibinfo {author}
  {\bibfnamefont {H.}~\bibnamefont {Song}}, \bibinfo {author} {\bibfnamefont
  {Z.-Z.}\ \bibnamefont {Yu}}, \ and\ \bibinfo {author} {\bibfnamefont
  {N.}~\bibnamefont {Koratkar}},\ }\href@noop {} {\bibfield  {journal}
  {\bibinfo  {journal} {ACS nano}\ }\textbf {\bibinfo {volume} {3}},\ \bibinfo
  {pages} {3884} (\bibinfo {year} {2009})}\BibitemShut {NoStop}%
\bibitem [{\citenamefont {Mesbah}\ \emph {et~al.}(2009)\citenamefont {Mesbah},
  \citenamefont {Za{\"\i}ri}, \citenamefont {Boutaleb}, \citenamefont
  {Gloaguen}, \citenamefont {Na{\"\i}t-Abdelaziz}, \citenamefont {Xie},
  \citenamefont {Boukharouba},\ and\ \citenamefont {Lefebvre}}]{Mesbah-2009}%
  \BibitemOpen
  \bibfield  {author} {\bibinfo {author} {\bibfnamefont {A.}~\bibnamefont
  {Mesbah}}, \bibinfo {author} {\bibfnamefont {F.}~\bibnamefont {Za{\"\i}ri}},
  \bibinfo {author} {\bibfnamefont {S.}~\bibnamefont {Boutaleb}}, \bibinfo
  {author} {\bibfnamefont {J.-M.}\ \bibnamefont {Gloaguen}}, \bibinfo {author}
  {\bibfnamefont {M.}~\bibnamefont {Na{\"\i}t-Abdelaziz}}, \bibinfo {author}
  {\bibfnamefont {S.}~\bibnamefont {Xie}}, \bibinfo {author} {\bibfnamefont
  {T.}~\bibnamefont {Boukharouba}}, \ and\ \bibinfo {author} {\bibfnamefont
  {J.-M.}\ \bibnamefont {Lefebvre}},\ }\href@noop {} {\bibfield  {journal}
  {\bibinfo  {journal} {Journal of applied polymer science}\ }\textbf {\bibinfo
  {volume} {114}},\ \bibinfo {pages} {3274} (\bibinfo {year}
  {2009})}\BibitemShut {NoStop}%
\bibitem [{\citenamefont {Fu}\ \emph {et~al.}(2008)\citenamefont {Fu},
  \citenamefont {Feng}, \citenamefont {Lauke},\ and\ \citenamefont
  {Mai}}]{Fu-2008}%
  \BibitemOpen
  \bibfield  {author} {\bibinfo {author} {\bibfnamefont {S.-Y.}\ \bibnamefont
  {Fu}}, \bibinfo {author} {\bibfnamefont {X.-Q.}\ \bibnamefont {Feng}},
  \bibinfo {author} {\bibfnamefont {B.}~\bibnamefont {Lauke}}, \ and\ \bibinfo
  {author} {\bibfnamefont {Y.-W.}\ \bibnamefont {Mai}},\ }\href@noop {}
  {\bibfield  {journal} {\bibinfo  {journal} {Composites Part B: Engineering}\
  }\textbf {\bibinfo {volume} {39}},\ \bibinfo {pages} {933} (\bibinfo {year}
  {2008})}\BibitemShut {NoStop}%
\bibitem [{\citenamefont {Ou}\ \emph {et~al.}(1998)\citenamefont {Ou},
  \citenamefont {Yang},\ and\ \citenamefont {Yu}}]{Ou-1998}%
  \BibitemOpen
  \bibfield  {author} {\bibinfo {author} {\bibfnamefont {Y.}~\bibnamefont
  {Ou}}, \bibinfo {author} {\bibfnamefont {F.}~\bibnamefont {Yang}}, \ and\
  \bibinfo {author} {\bibfnamefont {Z.-Z.}\ \bibnamefont {Yu}},\ }\href@noop {}
  {\bibfield  {journal} {\bibinfo  {journal} {Journal of Polymer Science Part
  B: Polymer Physics}\ }\textbf {\bibinfo {volume} {36}},\ \bibinfo {pages}
  {789} (\bibinfo {year} {1998})}\BibitemShut {NoStop}%
\bibitem [{\citenamefont {Wang}\ \emph {et~al.}(2002)\citenamefont {Wang},
  \citenamefont {Bai}, \citenamefont {Liu}, \citenamefont {Wu},\ and\
  \citenamefont {Wong}}]{Wang-2002}%
  \BibitemOpen
  \bibfield  {author} {\bibinfo {author} {\bibfnamefont {H.}~\bibnamefont
  {Wang}}, \bibinfo {author} {\bibfnamefont {Y.}~\bibnamefont {Bai}}, \bibinfo
  {author} {\bibfnamefont {S.}~\bibnamefont {Liu}}, \bibinfo {author}
  {\bibfnamefont {J.}~\bibnamefont {Wu}}, \ and\ \bibinfo {author}
  {\bibfnamefont {C.}~\bibnamefont {Wong}},\ }\href@noop {} {\bibfield
  {journal} {\bibinfo  {journal} {Acta materialia}\ }\textbf {\bibinfo {volume}
  {50}},\ \bibinfo {pages} {4369} (\bibinfo {year} {2002})}\BibitemShut
  {NoStop}%
\bibitem [{\citenamefont {Wetzel}\ \emph {et~al.}(2003)\citenamefont {Wetzel},
  \citenamefont {Haupert},\ and\ \citenamefont {Zhang}}]{Wetzel-2003}%
  \BibitemOpen
  \bibfield  {author} {\bibinfo {author} {\bibfnamefont {B.}~\bibnamefont
  {Wetzel}}, \bibinfo {author} {\bibfnamefont {F.}~\bibnamefont {Haupert}}, \
  and\ \bibinfo {author} {\bibfnamefont {M.~Q.}\ \bibnamefont {Zhang}},\
  }\href@noop {} {\bibfield  {journal} {\bibinfo  {journal} {Composites Science
  and Technology}\ }\textbf {\bibinfo {volume} {63}},\ \bibinfo {pages} {2055}
  (\bibinfo {year} {2003})}\BibitemShut {NoStop}%
\bibitem [{\citenamefont {Bershtein}\ \emph {et~al.}(2021)\citenamefont
  {Bershtein}, \citenamefont {Grigoryeva}, \citenamefont {Yakushev},\ and\
  \citenamefont {Fainleib}}]{Bershtein-2021}%
  \BibitemOpen
  \bibfield  {author} {\bibinfo {author} {\bibfnamefont {V.~A.}\ \bibnamefont
  {Bershtein}}, \bibinfo {author} {\bibfnamefont {O.~P.}\ \bibnamefont
  {Grigoryeva}}, \bibinfo {author} {\bibfnamefont {P.~N.}\ \bibnamefont
  {Yakushev}}, \ and\ \bibinfo {author} {\bibfnamefont {A.~M.}\ \bibnamefont
  {Fainleib}},\ }\href@noop {} {\bibfield  {journal} {\bibinfo  {journal}
  {Polymer Composites}\ }\textbf {\bibinfo {volume} {42}},\ \bibinfo {pages}
  {6777} (\bibinfo {year} {2021})}\BibitemShut {NoStop}%
\bibitem [{\citenamefont {Stojanovic}\ \emph {et~al.}(2009)\citenamefont
  {Stojanovic}, \citenamefont {Orlovic}, \citenamefont {Markovic},
  \citenamefont {Radmilovic}, \citenamefont {Uskokovic},\ and\ \citenamefont
  {Aleksic}}]{Stojanovic-2009}%
  \BibitemOpen
  \bibfield  {author} {\bibinfo {author} {\bibfnamefont {D.}~\bibnamefont
  {Stojanovic}}, \bibinfo {author} {\bibfnamefont {A.}~\bibnamefont {Orlovic}},
  \bibinfo {author} {\bibfnamefont {S.}~\bibnamefont {Markovic}}, \bibinfo
  {author} {\bibfnamefont {V.}~\bibnamefont {Radmilovic}}, \bibinfo {author}
  {\bibfnamefont {P.~S.}\ \bibnamefont {Uskokovic}}, \ and\ \bibinfo {author}
  {\bibfnamefont {R.}~\bibnamefont {Aleksic}},\ }\href@noop {} {\bibfield
  {journal} {\bibinfo  {journal} {Journal of Materials Science}\ }\textbf
  {\bibinfo {volume} {44}},\ \bibinfo {pages} {6223} (\bibinfo {year}
  {2009})}\BibitemShut {NoStop}%
\bibitem [{\citenamefont {Landau}\ \emph {et~al.}(1986)\citenamefont {Landau},
  \citenamefont {Lifshitz}, \citenamefont {Kosevich}, \citenamefont {Sykes},
  \citenamefont {Pitaevskii},\ and\ \citenamefont {Reid}}]{Landau-1986}%
  \BibitemOpen
  \bibfield  {author} {\bibinfo {author} {\bibfnamefont {L.}~\bibnamefont
  {Landau}}, \bibinfo {author} {\bibfnamefont {E.}~\bibnamefont {Lifshitz}},
  \bibinfo {author} {\bibfnamefont {A.}~\bibnamefont {Kosevich}}, \bibinfo
  {author} {\bibfnamefont {J.}~\bibnamefont {Sykes}}, \bibinfo {author}
  {\bibfnamefont {L.}~\bibnamefont {Pitaevskii}}, \ and\ \bibinfo {author}
  {\bibfnamefont {W.}~\bibnamefont {Reid}},\ }\href@noop {} {\emph {\bibinfo
  {title} {Theory of Elasticity: Volume 7}}},\ Course of theoretical physics\
  (\bibinfo  {publisher} {Elsevier Science},\ \bibinfo {year}
  {1986})\BibitemShut {NoStop}%
\bibitem [{\citenamefont {Eshelby}(1957)}]{Eshelby-1957}%
  \BibitemOpen
  \bibfield  {author} {\bibinfo {author} {\bibfnamefont {J.}~\bibnamefont
  {Eshelby}},\ }\href@noop {} {\bibfield  {journal} {\bibinfo  {journal}
  {Proceedings of the Royal Society of London. Series A. Mathematical and
  Physical Sciences}\ }\textbf {\bibinfo {volume} {241}},\ \bibinfo {pages}
  {376} (\bibinfo {year} {1957})}\BibitemShut {NoStop}%
\bibitem [{\citenamefont {Mori}\ and\ \citenamefont
  {Tanaka}(1973)}]{Mori-1973}%
  \BibitemOpen
  \bibfield  {author} {\bibinfo {author} {\bibfnamefont {T.}~\bibnamefont
  {Mori}}\ and\ \bibinfo {author} {\bibfnamefont {K.}~\bibnamefont {Tanaka}},\
  }\href@noop {} {\bibfield  {journal} {\bibinfo  {journal} {Acta
  metallurgica}\ }\textbf {\bibinfo {volume} {21}},\ \bibinfo {pages} {571}
  (\bibinfo {year} {1973})}\BibitemShut {NoStop}%
\bibitem [{\citenamefont {Benveniste}(1987)}]{Benveniste-1987}%
  \BibitemOpen
  \bibfield  {author} {\bibinfo {author} {\bibfnamefont {Y.}~\bibnamefont
  {Benveniste}},\ }\href@noop {} {\bibfield  {journal} {\bibinfo  {journal}
  {Mechanics of materials}\ }\textbf {\bibinfo {volume} {6}},\ \bibinfo {pages}
  {147} (\bibinfo {year} {1987})}\BibitemShut {NoStop}%
\bibitem [{\citenamefont {Odegard}\ \emph {et~al.}(2005)\citenamefont
  {Odegard}, \citenamefont {Clancy},\ and\ \citenamefont
  {Gates}}]{Odegard-2005}%
  \BibitemOpen
  \bibfield  {author} {\bibinfo {author} {\bibfnamefont {G.}~\bibnamefont
  {Odegard}}, \bibinfo {author} {\bibfnamefont {T.}~\bibnamefont {Clancy}}, \
  and\ \bibinfo {author} {\bibfnamefont {T.}~\bibnamefont {Gates}},\
  }\href@noop {} {\bibfield  {journal} {\bibinfo  {journal} {Polymer}\ }\textbf
  {\bibinfo {volume} {46}},\ \bibinfo {pages} {553} (\bibinfo {year}
  {2005})}\BibitemShut {NoStop}%
\bibitem [{\citenamefont {Bondioli}\ \emph {et~al.}(2005)\citenamefont
  {Bondioli}, \citenamefont {Cannillo}, \citenamefont {Fabbri},\ and\
  \citenamefont {Messori}}]{Bondioli-2005}%
  \BibitemOpen
  \bibfield  {author} {\bibinfo {author} {\bibfnamefont {F.}~\bibnamefont
  {Bondioli}}, \bibinfo {author} {\bibfnamefont {V.}~\bibnamefont {Cannillo}},
  \bibinfo {author} {\bibfnamefont {E.}~\bibnamefont {Fabbri}}, \ and\ \bibinfo
  {author} {\bibfnamefont {M.}~\bibnamefont {Messori}},\ }\href@noop {}
  {\bibfield  {journal} {\bibinfo  {journal} {Journal of applied polymer
  science}\ }\textbf {\bibinfo {volume} {97}},\ \bibinfo {pages} {2382}
  (\bibinfo {year} {2005})}\BibitemShut {NoStop}%
\bibitem [{\citenamefont {Saber-Samandari}\ and\ \citenamefont
  {Afaghi-Khatibi}(2007)}]{Saber-2007}%
  \BibitemOpen
  \bibfield  {author} {\bibinfo {author} {\bibfnamefont {S.}~\bibnamefont
  {Saber-Samandari}}\ and\ \bibinfo {author} {\bibfnamefont {A.}~\bibnamefont
  {Afaghi-Khatibi}},\ }\href@noop {} {\bibfield  {journal} {\bibinfo  {journal}
  {Polymer composites}\ }\textbf {\bibinfo {volume} {28}},\ \bibinfo {pages}
  {405} (\bibinfo {year} {2007})}\BibitemShut {NoStop}%
\bibitem [{\citenamefont {Qiao}\ and\ \citenamefont
  {Brinson}(2009)}]{Qiao-2009}%
  \BibitemOpen
  \bibfield  {author} {\bibinfo {author} {\bibfnamefont {R.}~\bibnamefont
  {Qiao}}\ and\ \bibinfo {author} {\bibfnamefont {L.~C.}\ \bibnamefont
  {Brinson}},\ }\href@noop {} {\bibfield  {journal} {\bibinfo  {journal}
  {Composites Science and Technology}\ }\textbf {\bibinfo {volume} {69}},\
  \bibinfo {pages} {491} (\bibinfo {year} {2009})}\BibitemShut {NoStop}%
\bibitem [{\citenamefont {Wang}\ \emph {et~al.}(2011)\citenamefont {Wang},
  \citenamefont {Zhou}, \citenamefont {Peng},\ and\ \citenamefont
  {Mishnaevsky~Jr}}]{Wang-2011}%
  \BibitemOpen
  \bibfield  {author} {\bibinfo {author} {\bibfnamefont {H.}~\bibnamefont
  {Wang}}, \bibinfo {author} {\bibfnamefont {H.}~\bibnamefont {Zhou}}, \bibinfo
  {author} {\bibfnamefont {R.}~\bibnamefont {Peng}}, \ and\ \bibinfo {author}
  {\bibfnamefont {L.}~\bibnamefont {Mishnaevsky~Jr}},\ }\href@noop {}
  {\bibfield  {journal} {\bibinfo  {journal} {Composites Science and
  Technology}\ }\textbf {\bibinfo {volume} {71}},\ \bibinfo {pages} {980}
  (\bibinfo {year} {2011})}\BibitemShut {NoStop}%
\bibitem [{\citenamefont {Amraei}\ \emph {et~al.}(2019)\citenamefont {Amraei},
  \citenamefont {Jam}, \citenamefont {Arab},\ and\ \citenamefont
  {Firouz-Abadi}}]{Amraei-2019}%
  \BibitemOpen
  \bibfield  {author} {\bibinfo {author} {\bibfnamefont {J.}~\bibnamefont
  {Amraei}}, \bibinfo {author} {\bibfnamefont {J.~E.}\ \bibnamefont {Jam}},
  \bibinfo {author} {\bibfnamefont {B.}~\bibnamefont {Arab}}, \ and\ \bibinfo
  {author} {\bibfnamefont {R.~D.}\ \bibnamefont {Firouz-Abadi}},\ }\href@noop
  {} {\bibfield  {journal} {\bibinfo  {journal} {Journal of Composite
  Materials}\ }\textbf {\bibinfo {volume} {53}},\ \bibinfo {pages} {1261}
  (\bibinfo {year} {2019})}\BibitemShut {NoStop}%
\bibitem [{\citenamefont {Beltukov}\ \emph {et~al.}(2013)\citenamefont
  {Beltukov}, \citenamefont {Kozub},\ and\ \citenamefont
  {Parshin}}]{Beltukov-2013}%
  \BibitemOpen
  \bibfield  {author} {\bibinfo {author} {\bibfnamefont {Y.}~\bibnamefont
  {Beltukov}}, \bibinfo {author} {\bibfnamefont {V.}~\bibnamefont {Kozub}}, \
  and\ \bibinfo {author} {\bibfnamefont {D.}~\bibnamefont {Parshin}},\
  }\href@noop {} {\bibfield  {journal} {\bibinfo  {journal} {Physical Review
  B}\ }\textbf {\bibinfo {volume} {87}},\ \bibinfo {pages} {134203} (\bibinfo
  {year} {2013})}\BibitemShut {NoStop}%
\bibitem [{\citenamefont {Jana}\ and\ \citenamefont
  {Pastewka}(2019)}]{Jana-2019}%
  \BibitemOpen
  \bibfield  {author} {\bibinfo {author} {\bibfnamefont {R.}~\bibnamefont
  {Jana}}\ and\ \bibinfo {author} {\bibfnamefont {L.}~\bibnamefont
  {Pastewka}},\ }\href@noop {} {\bibfield  {journal} {\bibinfo  {journal}
  {Journal of Physics: Materials}\ }\textbf {\bibinfo {volume} {2}},\ \bibinfo
  {pages} {045006} (\bibinfo {year} {2019})}\BibitemShut {NoStop}%
\bibitem [{\citenamefont {Wen}\ \emph {et~al.}(2012)\citenamefont {Wen},
  \citenamefont {Basu}, \citenamefont {Janmey},\ and\ \citenamefont
  {Yodh}}]{Wen-2012}%
  \BibitemOpen
  \bibfield  {author} {\bibinfo {author} {\bibfnamefont {Q.}~\bibnamefont
  {Wen}}, \bibinfo {author} {\bibfnamefont {A.}~\bibnamefont {Basu}}, \bibinfo
  {author} {\bibfnamefont {P.~A.}\ \bibnamefont {Janmey}}, \ and\ \bibinfo
  {author} {\bibfnamefont {A.~G.}\ \bibnamefont {Yodh}},\ }\href@noop {}
  {\bibfield  {journal} {\bibinfo  {journal} {Soft matter}\ }\textbf {\bibinfo
  {volume} {8}},\ \bibinfo {pages} {8039} (\bibinfo {year} {2012})}\BibitemShut
  {NoStop}%
\bibitem [{\citenamefont {Del~Gado}\ \emph {et~al.}(2008)\citenamefont
  {Del~Gado}, \citenamefont {Ilg}, \citenamefont {Kr{\"o}ger},\ and\
  \citenamefont {{\"O}ttinger}}]{Del-2008}%
  \BibitemOpen
  \bibfield  {author} {\bibinfo {author} {\bibfnamefont {E.}~\bibnamefont
  {Del~Gado}}, \bibinfo {author} {\bibfnamefont {P.}~\bibnamefont {Ilg}},
  \bibinfo {author} {\bibfnamefont {M.}~\bibnamefont {Kr{\"o}ger}}, \ and\
  \bibinfo {author} {\bibfnamefont {H.~C.}\ \bibnamefont {{\"O}ttinger}},\
  }\href@noop {} {\bibfield  {journal} {\bibinfo  {journal} {Physical review
  letters}\ }\textbf {\bibinfo {volume} {101}},\ \bibinfo {pages} {095501}
  (\bibinfo {year} {2008})}\BibitemShut {NoStop}%
\bibitem [{\citenamefont {Leonforte}\ \emph {et~al.}(2006)\citenamefont
  {Leonforte}, \citenamefont {Tanguy}, \citenamefont {Wittmer},\ and\
  \citenamefont {Barrat}}]{Leonforte-2006}%
  \BibitemOpen
  \bibfield  {author} {\bibinfo {author} {\bibfnamefont {F.}~\bibnamefont
  {Leonforte}}, \bibinfo {author} {\bibfnamefont {A.}~\bibnamefont {Tanguy}},
  \bibinfo {author} {\bibfnamefont {J.}~\bibnamefont {Wittmer}}, \ and\
  \bibinfo {author} {\bibfnamefont {J.-L.}\ \bibnamefont {Barrat}},\
  }\href@noop {} {\bibfield  {journal} {\bibinfo  {journal} {Physical review
  letters}\ }\textbf {\bibinfo {volume} {97}},\ \bibinfo {pages} {055501}
  (\bibinfo {year} {2006})}\BibitemShut {NoStop}%
\bibitem [{\citenamefont {Leonforte}\ \emph {et~al.}(2005)\citenamefont
  {Leonforte}, \citenamefont {Boissi{\`e}re}, \citenamefont {Tanguy},
  \citenamefont {Wittmer},\ and\ \citenamefont {Barrat}}]{Leonforte-2005}%
  \BibitemOpen
  \bibfield  {author} {\bibinfo {author} {\bibfnamefont {F.}~\bibnamefont
  {Leonforte}}, \bibinfo {author} {\bibfnamefont {R.}~\bibnamefont
  {Boissi{\`e}re}}, \bibinfo {author} {\bibfnamefont {A.}~\bibnamefont
  {Tanguy}}, \bibinfo {author} {\bibfnamefont {J.}~\bibnamefont {Wittmer}}, \
  and\ \bibinfo {author} {\bibfnamefont {J.-L.}\ \bibnamefont {Barrat}},\
  }\href@noop {} {\bibfield  {journal} {\bibinfo  {journal} {Physical Review
  B}\ }\textbf {\bibinfo {volume} {72}},\ \bibinfo {pages} {224206} (\bibinfo
  {year} {2005})}\BibitemShut {NoStop}%
\bibitem [{\citenamefont {Conyuh}\ \emph {et~al.}(2019)\citenamefont {Conyuh},
  \citenamefont {Beltukov},\ and\ \citenamefont {Parshin}}]{Conyuh-2019}%
  \BibitemOpen
  \bibfield  {author} {\bibinfo {author} {\bibfnamefont {D.}~\bibnamefont
  {Conyuh}}, \bibinfo {author} {\bibfnamefont {Y.}~\bibnamefont {Beltukov}}, \
  and\ \bibinfo {author} {\bibfnamefont {D.}~\bibnamefont {Parshin}},\
  }\href@noop {} {\bibfield  {journal} {\bibinfo  {journal} {Physics of the
  Solid State}\ }\textbf {\bibinfo {volume} {61}},\ \bibinfo {pages} {1272}
  (\bibinfo {year} {2019})}\BibitemShut {NoStop}%
\bibitem [{\citenamefont {Marrink}\ \emph {et~al.}(2007)\citenamefont
  {Marrink}, \citenamefont {Risselada}, \citenamefont {Yefimov}, \citenamefont
  {Tieleman},\ and\ \citenamefont {de~Vries}}]{Marrink2007}%
  \BibitemOpen
  \bibfield  {author} {\bibinfo {author} {\bibfnamefont {S.}~\bibnamefont
  {Marrink}}, \bibinfo {author} {\bibfnamefont {H.}~\bibnamefont {Risselada}},
  \bibinfo {author} {\bibfnamefont {S.}~\bibnamefont {Yefimov}}, \bibinfo
  {author} {\bibfnamefont {D.}~\bibnamefont {Tieleman}}, \ and\ \bibinfo
  {author} {\bibfnamefont {A.}~\bibnamefont {de~Vries}},\ }\href@noop {}
  {\bibfield  {journal} {\bibinfo  {journal} {J. Phys. Chem. B}\ }\textbf
  {\bibinfo {volume} {111}},\ \bibinfo {pages} {7812} (\bibinfo {year}
  {2007})}\BibitemShut {NoStop}%
\bibitem [{\citenamefont {Rossi}\ \emph {et~al.}(2011)\citenamefont {Rossi},
  \citenamefont {Monticelli}, \citenamefont {Puisto}, \citenamefont
  {Vattulainen},\ and\ \citenamefont {Ala-Nissila}}]{rossi2011coarse}%
  \BibitemOpen
  \bibfield  {author} {\bibinfo {author} {\bibfnamefont {G.}~\bibnamefont
  {Rossi}}, \bibinfo {author} {\bibfnamefont {L.}~\bibnamefont {Monticelli}},
  \bibinfo {author} {\bibfnamefont {S.~R.}\ \bibnamefont {Puisto}}, \bibinfo
  {author} {\bibfnamefont {I.}~\bibnamefont {Vattulainen}}, \ and\ \bibinfo
  {author} {\bibfnamefont {T.}~\bibnamefont {Ala-Nissila}},\ }\href@noop {}
  {\bibfield  {journal} {\bibinfo  {journal} {Soft Matter}\ }\textbf {\bibinfo
  {volume} {7}},\ \bibinfo {pages} {698} (\bibinfo {year} {2011})}\BibitemShut
  {NoStop}%
\bibitem [{\citenamefont {Beltukov}\ \emph {et~al.}(2019)\citenamefont
  {Beltukov}, \citenamefont {Gula}, \citenamefont {Samsonov},\ and\
  \citenamefont {Solov'yov}}]{Beltukov2019}%
  \BibitemOpen
  \bibfield  {author} {\bibinfo {author} {\bibfnamefont {Y.~M.}\ \bibnamefont
  {Beltukov}}, \bibinfo {author} {\bibfnamefont {I.}~\bibnamefont {Gula}},
  \bibinfo {author} {\bibfnamefont {A.~M.}\ \bibnamefont {Samsonov}}, \ and\
  \bibinfo {author} {\bibfnamefont {I.~A.}\ \bibnamefont {Solov'yov}},\
  }\href@noop {} {\bibfield  {journal} {\bibinfo  {journal} {Eur.\ Phys.\ J.\
  D}\ }\textbf {\bibinfo {volume} {73}},\ \bibinfo {pages} {226} (\bibinfo
  {year} {2019})}\BibitemShut {NoStop}%
\bibitem [{\citenamefont {Phillips}\ \emph {et~al.}(2005)\citenamefont
  {Phillips}, \citenamefont {Braun}, \citenamefont {Wang}, \citenamefont
  {Gumbart}, \citenamefont {Tajkhorshid}, \citenamefont {Villa}, \citenamefont
  {Chipot}, \citenamefont {Skeel}, \citenamefont {Kale},\ and\ \citenamefont
  {Schulten}}]{PHIL2005}%
  \BibitemOpen
  \bibfield  {author} {\bibinfo {author} {\bibfnamefont {J.~C.}\ \bibnamefont
  {Phillips}}, \bibinfo {author} {\bibfnamefont {R.}~\bibnamefont {Braun}},
  \bibinfo {author} {\bibfnamefont {W.}~\bibnamefont {Wang}}, \bibinfo {author}
  {\bibfnamefont {J.}~\bibnamefont {Gumbart}}, \bibinfo {author} {\bibfnamefont
  {E.}~\bibnamefont {Tajkhorshid}}, \bibinfo {author} {\bibfnamefont
  {E.}~\bibnamefont {Villa}}, \bibinfo {author} {\bibfnamefont
  {C.}~\bibnamefont {Chipot}}, \bibinfo {author} {\bibfnamefont {R.~D.}\
  \bibnamefont {Skeel}}, \bibinfo {author} {\bibfnamefont {L.}~\bibnamefont
  {Kale}}, \ and\ \bibinfo {author} {\bibfnamefont {K.}~\bibnamefont
  {Schulten}},\ }\href@noop {} {\bibfield  {journal} {\bibinfo  {journal}
  {J.~Comp.\ Chem.}\ }\textbf {\bibinfo {volume} {26}},\ \bibinfo {pages}
  {1781} (\bibinfo {year} {2005})}\BibitemShut {NoStop}%
\bibitem [{\citenamefont {Humphrey}\ \emph {et~al.}(1996)\citenamefont
  {Humphrey}, \citenamefont {Dalke},\ and\ \citenamefont {Schulten}}]{HUMP96}%
  \BibitemOpen
  \bibfield  {author} {\bibinfo {author} {\bibfnamefont {W.}~\bibnamefont
  {Humphrey}}, \bibinfo {author} {\bibfnamefont {A.}~\bibnamefont {Dalke}}, \
  and\ \bibinfo {author} {\bibfnamefont {K.}~\bibnamefont {Schulten}},\
  }\href@noop {} {\bibfield  {journal} {\bibinfo  {journal} {J. Molec.
  Graphics}\ }\textbf {\bibinfo {volume} {14}},\ \bibinfo {pages} {33}
  (\bibinfo {year} {1996})}\BibitemShut {NoStop}%
\bibitem [{\citenamefont {Sushko}\ \emph {et~al.}(2019)\citenamefont {Sushko},
  \citenamefont {Solov'yov},\ and\ \citenamefont {Solov'yov}}]{Sushko2019}%
  \BibitemOpen
  \bibfield  {author} {\bibinfo {author} {\bibfnamefont {G.~B.}\ \bibnamefont
  {Sushko}}, \bibinfo {author} {\bibfnamefont {I.~A.}\ \bibnamefont
  {Solov'yov}}, \ and\ \bibinfo {author} {\bibfnamefont {A.~V.}\ \bibnamefont
  {Solov'yov}},\ }\href@noop {} {\bibfield  {journal} {\bibinfo  {journal}
  {J.~Mol.\ Graphics and Modelling}\ }\textbf {\bibinfo {volume} {88}},\
  \bibinfo {pages} {247} (\bibinfo {year} {2019})}\BibitemShut {NoStop}%
\bibitem [{\citenamefont {Vollmayr}\ \emph {et~al.}(1996)\citenamefont
  {Vollmayr}, \citenamefont {Kob},\ and\ \citenamefont
  {Binder}}]{Vollmayr1996}%
  \BibitemOpen
  \bibfield  {author} {\bibinfo {author} {\bibfnamefont {K.}~\bibnamefont
  {Vollmayr}}, \bibinfo {author} {\bibfnamefont {W.}~\bibnamefont {Kob}}, \
  and\ \bibinfo {author} {\bibfnamefont {K.}~\bibnamefont {Binder}},\
  }\href@noop {} {\bibfield  {journal} {\bibinfo  {journal} {Physical Review
  B}\ }\textbf {\bibinfo {volume} {54}},\ \bibinfo {pages} {15808} (\bibinfo
  {year} {1996})}\BibitemShut {NoStop}%
\bibitem [{\citenamefont {Sengupta}\ \emph {et~al.}(2000)\citenamefont
  {Sengupta}, \citenamefont {Nielaba}, \citenamefont {Rao},\ and\ \citenamefont
  {Binder}}]{Sengupta-2000}%
  \BibitemOpen
  \bibfield  {author} {\bibinfo {author} {\bibfnamefont {S.}~\bibnamefont
  {Sengupta}}, \bibinfo {author} {\bibfnamefont {P.}~\bibnamefont {Nielaba}},
  \bibinfo {author} {\bibfnamefont {M.}~\bibnamefont {Rao}}, \ and\ \bibinfo
  {author} {\bibfnamefont {K.}~\bibnamefont {Binder}},\ }\href@noop {}
  {\bibfield  {journal} {\bibinfo  {journal} {Physical Review E}\ }\textbf
  {\bibinfo {volume} {61}},\ \bibinfo {pages} {1072} (\bibinfo {year}
  {2000})}\BibitemShut {NoStop}%
\bibitem [{\citenamefont {Yoshimoto}\ \emph {et~al.}(2004)\citenamefont
  {Yoshimoto}, \citenamefont {Jain}, \citenamefont {Van~Workum}, \citenamefont
  {Nealey},\ and\ \citenamefont {de~Pablo}}]{Yoshimoto-2004}%
  \BibitemOpen
  \bibfield  {author} {\bibinfo {author} {\bibfnamefont {K.}~\bibnamefont
  {Yoshimoto}}, \bibinfo {author} {\bibfnamefont {T.~S.}\ \bibnamefont {Jain}},
  \bibinfo {author} {\bibfnamefont {K.}~\bibnamefont {Van~Workum}}, \bibinfo
  {author} {\bibfnamefont {P.~F.}\ \bibnamefont {Nealey}}, \ and\ \bibinfo
  {author} {\bibfnamefont {J.~J.}\ \bibnamefont {de~Pablo}},\ }\href@noop {}
  {\bibfield  {journal} {\bibinfo  {journal} {Physical review letters}\
  }\textbf {\bibinfo {volume} {93}},\ \bibinfo {pages} {175501} (\bibinfo
  {year} {2004})}\BibitemShut {NoStop}%
\bibitem [{\citenamefont {Tsamados}\ \emph {et~al.}(2009)\citenamefont
  {Tsamados}, \citenamefont {Tanguy}, \citenamefont {Goldenberg},\ and\
  \citenamefont {Barrat}}]{Tsamados-2009}%
  \BibitemOpen
  \bibfield  {author} {\bibinfo {author} {\bibfnamefont {M.}~\bibnamefont
  {Tsamados}}, \bibinfo {author} {\bibfnamefont {A.}~\bibnamefont {Tanguy}},
  \bibinfo {author} {\bibfnamefont {C.}~\bibnamefont {Goldenberg}}, \ and\
  \bibinfo {author} {\bibfnamefont {J.-L.}\ \bibnamefont {Barrat}},\
  }\href@noop {} {\bibfield  {journal} {\bibinfo  {journal} {Physical Review
  E}\ }\textbf {\bibinfo {volume} {80}},\ \bibinfo {pages} {026112} (\bibinfo
  {year} {2009})}\BibitemShut {NoStop}%
\bibitem [{\citenamefont {Mizuno}\ \emph {et~al.}(2013)\citenamefont {Mizuno},
  \citenamefont {Mossa},\ and\ \citenamefont {Barrat}}]{Mizuno-2013}%
  \BibitemOpen
  \bibfield  {author} {\bibinfo {author} {\bibfnamefont {H.}~\bibnamefont
  {Mizuno}}, \bibinfo {author} {\bibfnamefont {S.}~\bibnamefont {Mossa}}, \
  and\ \bibinfo {author} {\bibfnamefont {J.-L.}\ \bibnamefont {Barrat}},\
  }\href@noop {} {\bibfield  {journal} {\bibinfo  {journal} {Physical Review
  E}\ }\textbf {\bibinfo {volume} {87}},\ \bibinfo {pages} {042306} (\bibinfo
  {year} {2013})}\BibitemShut {NoStop}%
\bibitem [{\citenamefont {Sussman}\ \emph {et~al.}(2015)\citenamefont
  {Sussman}, \citenamefont {Schoenholz}, \citenamefont {Xu}, \citenamefont
  {Still}, \citenamefont {Yodh},\ and\ \citenamefont {Liu}}]{Sussman-2015}%
  \BibitemOpen
  \bibfield  {author} {\bibinfo {author} {\bibfnamefont {D.~M.}\ \bibnamefont
  {Sussman}}, \bibinfo {author} {\bibfnamefont {S.~S.}\ \bibnamefont
  {Schoenholz}}, \bibinfo {author} {\bibfnamefont {Y.}~\bibnamefont {Xu}},
  \bibinfo {author} {\bibfnamefont {T.}~\bibnamefont {Still}}, \bibinfo
  {author} {\bibfnamefont {A.}~\bibnamefont {Yodh}}, \ and\ \bibinfo {author}
  {\bibfnamefont {A.~J.}\ \bibnamefont {Liu}},\ }\href@noop {} {\bibfield
  {journal} {\bibinfo  {journal} {Physical Review E}\ }\textbf {\bibinfo
  {volume} {92}},\ \bibinfo {pages} {022307} (\bibinfo {year}
  {2015})}\BibitemShut {NoStop}%
\bibitem [{\citenamefont {Amuasi}\ \emph {et~al.}(2015)\citenamefont {Amuasi},
  \citenamefont {Heussinger}, \citenamefont {Vink},\ and\ \citenamefont
  {Zippelius}}]{Amuasi-2015}%
  \BibitemOpen
  \bibfield  {author} {\bibinfo {author} {\bibfnamefont {H.}~\bibnamefont
  {Amuasi}}, \bibinfo {author} {\bibfnamefont {C.}~\bibnamefont {Heussinger}},
  \bibinfo {author} {\bibfnamefont {R.}~\bibnamefont {Vink}}, \ and\ \bibinfo
  {author} {\bibfnamefont {A.}~\bibnamefont {Zippelius}},\ }\href@noop {}
  {\bibfield  {journal} {\bibinfo  {journal} {New Journal of Physics}\ }\textbf
  {\bibinfo {volume} {17}},\ \bibinfo {pages} {083035} (\bibinfo {year}
  {2015})}\BibitemShut {NoStop}%
\bibitem [{\citenamefont {Mavko}\ \emph {et~al.}(1995)\citenamefont {Mavko},
  \citenamefont {Chan},\ and\ \citenamefont {Mukerji}}]{Mavko-1995}%
  \BibitemOpen
  \bibfield  {author} {\bibinfo {author} {\bibfnamefont {G.}~\bibnamefont
  {Mavko}}, \bibinfo {author} {\bibfnamefont {C.}~\bibnamefont {Chan}}, \ and\
  \bibinfo {author} {\bibfnamefont {T.}~\bibnamefont {Mukerji}},\ }\href@noop
  {} {\bibfield  {journal} {\bibinfo  {journal} {Geophysics}\ }\textbf
  {\bibinfo {volume} {60}},\ \bibinfo {pages} {1750} (\bibinfo {year}
  {1995})}\BibitemShut {NoStop}%
\bibitem [{\citenamefont {Mavko}\ \emph {et~al.}(2003)\citenamefont {Mavko},
  \citenamefont {Mukerji},\ and\ \citenamefont {Dvorkin}}]{Mavko-2003}%
  \BibitemOpen
  \bibfield  {author} {\bibinfo {author} {\bibfnamefont {G.}~\bibnamefont
  {Mavko}}, \bibinfo {author} {\bibfnamefont {T.}~\bibnamefont {Mukerji}}, \
  and\ \bibinfo {author} {\bibfnamefont {J.}~\bibnamefont {Dvorkin}},\
  }\href@noop {} {\emph {\bibinfo {title} {The Rock Physics Handbook: Tools for
  Seismic Analysis of Porous Media}}},\ Stanford-Cambridge program\ (\bibinfo
  {publisher} {Cambridge University Press},\ \bibinfo {year}
  {2003})\BibitemShut {NoStop}%
\bibitem [{\citenamefont {Gilmour}\ \emph {et~al.}(1979)\citenamefont
  {Gilmour}, \citenamefont {Trainor},\ and\ \citenamefont
  {Haward}}]{Gilmour-1979}%
  \BibitemOpen
  \bibfield  {author} {\bibinfo {author} {\bibfnamefont {I.}~\bibnamefont
  {Gilmour}}, \bibinfo {author} {\bibfnamefont {A.}~\bibnamefont {Trainor}}, \
  and\ \bibinfo {author} {\bibfnamefont {R.}~\bibnamefont {Haward}},\
  }\href@noop {} {\bibfield  {journal} {\bibinfo  {journal} {Journal of Applied
  Polymer Science}\ }\textbf {\bibinfo {volume} {23}},\ \bibinfo {pages} {3129}
  (\bibinfo {year} {1979})}\BibitemShut {NoStop}%
\bibitem [{\citenamefont {Deschamps}\ \emph {et~al.}(2014)\citenamefont
  {Deschamps}, \citenamefont {Margueritat}, \citenamefont {Martinet},
  \citenamefont {Mermet},\ and\ \citenamefont {Champagnon}}]{Deschamps-2014}%
  \BibitemOpen
  \bibfield  {author} {\bibinfo {author} {\bibfnamefont {T.}~\bibnamefont
  {Deschamps}}, \bibinfo {author} {\bibfnamefont {J.}~\bibnamefont
  {Margueritat}}, \bibinfo {author} {\bibfnamefont {C.}~\bibnamefont
  {Martinet}}, \bibinfo {author} {\bibfnamefont {A.}~\bibnamefont {Mermet}}, \
  and\ \bibinfo {author} {\bibfnamefont {B.}~\bibnamefont {Champagnon}},\
  }\href@noop {} {\bibfield  {journal} {\bibinfo  {journal} {Scientific
  reports}\ }\textbf {\bibinfo {volume} {4}},\ \bibinfo {pages} {1} (\bibinfo
  {year} {2014})}\BibitemShut {NoStop}%
\bibitem [{\citenamefont {Beltukov}\ and\ \citenamefont
  {Skipetrov}(2017)}]{Beltukov-2017}%
  \BibitemOpen
  \bibfield  {author} {\bibinfo {author} {\bibfnamefont {Y.}~\bibnamefont
  {Beltukov}}\ and\ \bibinfo {author} {\bibfnamefont {S.}~\bibnamefont
  {Skipetrov}},\ }\href@noop {} {\bibfield  {journal} {\bibinfo  {journal}
  {Physical Review B}\ }\textbf {\bibinfo {volume} {96}},\ \bibinfo {pages}
  {174209} (\bibinfo {year} {2017})}\BibitemShut {NoStop}%
\bibitem [{\citenamefont {Conyuh}\ and\ \citenamefont
  {Beltukov}(2021)}]{Conyuh-2021}%
  \BibitemOpen
  \bibfield  {author} {\bibinfo {author} {\bibfnamefont {D.}~\bibnamefont
  {Conyuh}}\ and\ \bibinfo {author} {\bibfnamefont {Y.}~\bibnamefont
  {Beltukov}},\ }\href@noop {} {\bibfield  {journal} {\bibinfo  {journal}
  {Physical Review B}\ }\textbf {\bibinfo {volume} {103}},\ \bibinfo {pages}
  {104204} (\bibinfo {year} {2021})}\BibitemShut {NoStop}%
\bibitem [{\citenamefont {Fankh{\"a}nel}\ \emph {et~al.}(2019)\citenamefont
  {Fankh{\"a}nel}, \citenamefont {Arash},\ and\ \citenamefont
  {Rolfes}}]{Fankhanel-2019}%
  \BibitemOpen
  \bibfield  {author} {\bibinfo {author} {\bibfnamefont {J.}~\bibnamefont
  {Fankh{\"a}nel}}, \bibinfo {author} {\bibfnamefont {B.}~\bibnamefont
  {Arash}}, \ and\ \bibinfo {author} {\bibfnamefont {R.}~\bibnamefont
  {Rolfes}},\ }\href@noop {} {\bibfield  {journal} {\bibinfo  {journal}
  {Composites Part B: Engineering}\ }\textbf {\bibinfo {volume} {176}},\
  \bibinfo {pages} {107211} (\bibinfo {year} {2019})}\BibitemShut {NoStop}%
\bibitem [{\citenamefont {Gr{\o}nbech-Jensen}(2020)}]{Gronbech-2020}%
  \BibitemOpen
  \bibfield  {author} {\bibinfo {author} {\bibfnamefont {N.}~\bibnamefont
  {Gr{\o}nbech-Jensen}},\ }\href@noop {} {\bibfield  {journal} {\bibinfo
  {journal} {Molecular Physics}\ }\textbf {\bibinfo {volume} {118}},\ \bibinfo
  {pages} {e1662506} (\bibinfo {year} {2020})}\BibitemShut {NoStop}%
\end{thebibliography}%

\newpage

\newcommand{\Tr}{\operatorname{Tr}}

\renewcommand\theequation{S\arabic{equation}}
\renewcommand\thepage{S\arabic{page}}
\renewcommand\thefigure{S\arabic{figure}}

\section*{\hspace{4cm}Supplemental Material}
\label{sm}

\section{Density fluctuations and local elasticity modulus}
\label{sec:fluct}

Let us consider an elastic cubic sample with the dimensions of $L\times L\times L$ with periodic boundaries in the thermal equilibrium at a temperature $T$. Thermal fluctuations can be described by a displacement field
\begin{equation}
{\bf u}({\bf r}) = {\sum_{\bf q}}'\big[{\bf a}_{\bf q} \sin({\bf q}\cdot{\bf r}) + {\bf b}_{\bf q} \cos({\bf q}\cdot{\bf r})\big],  \label{eq:u}
\end{equation}
where the wavevector ${\bf q}$ is determined by the three integer numbers $(n_x, n_y, n_z)$ as $q_i = 2\pi n_i/L$ for $i=x,y,z$. Wavevectors ${\bf q}$ and $-{\bf q}$ define the same wave with ${\bf a}_{-{\bf q}} = -{\bf a}_{\bf q}$ and ${\bf b}_{-{\bf q}} = {\bf b}_{\bf q}$. The summation in Eq.~(\ref{eq:u}) is performed over half of the reciprocal space with non-equivalent wavevectors, which is denoted by a prime. The wavevector ${\bf q} = 0$ is excluded from the summation as it corresponds to a trivial translation of the system.

The deviation of the local density is defined as
\begin{equation}
\delta\rho({\bf r}_0) = \rho_0 \int \phi({\bf r} - {\bf r}_0)\Tr{\hat{\varepsilon}({\bf r})} d{\bf r},
\end{equation}
where $\rho_0$ is the equilibrium density, $\phi({\bf r})$ is the smoothing function defined in Eq.~(\ref{eq:phi}), and $\hat{\varepsilon}({\bf r})$ is the strain tensor written as ~\cite{Landau-1986}
\begin{eqnarray}
\nonumber
\varepsilon_{ij} &=& \frac{1}{2}\left(\frac{\partial u_i}{\partial x_j} + \frac{\partial u_j}{\partial x_i}\right) \\
\nonumber
&=& \frac{1}{2}{\sum_{\bf q}}'\Big[(a_{{\bf q}i}q_j + a_{{\bf q}j}q_i) \cos({\bf q}\cdot{\bf r}) \\
&&- (b_{{\bf q}i}q_j + b_{{\bf q}j}q_i) \sin({\bf q}\cdot{\bf r})\Big].
\end{eqnarray}
As a result, the deviation of the local density is $\delta\rho = \sum_{\bf q}' \delta\rho_{\bf q}$, where
\begin{eqnarray}
\nonumber
\delta\rho_{\bf q} &=& \rho_0 \big[({\bf a_q}\cdot{\bf q})\cos({\bf q}\cdot {\bf r}_0) \\
&&+ ({\bf b_q}\cdot{\bf q})\sin({\bf q}\cdot {\bf r}_0)  \big] \exp\left(-\frac{w^2q^2}{2}\right).
\end{eqnarray}

To find the relation between the local density fluctuations and elastic moduli, consider a homogeneous isotropic elastic body as a reference. In that case, the elastic energy density is
\begin{equation}
\Pi({\bf r}) = \frac{1}{2} C_{ijkl} \varepsilon_{ij}({\bf r})\varepsilon_{kl}({\bf r}),
\end{equation}
where the stiffness tensor is defined as
\begin{equation}
C_{ijkl} = \lambda\delta_{ij}\delta_{kl} + \mu (\delta_{ik}\delta_{jl} + \delta_{il}\delta_{jk}),
\end{equation}
with $\lambda$ and $\mu$ being the Lam\'e parameters~\cite{Landau-1986}. The total elastic energy of the system is given as $U = \int \Pi({\bf r})d{\bf r} = \sum_{\bf q}' U_{\bf q}$, where
\begin{eqnarray}
\nonumber
U_{\bf q} &=&  \frac{L^3}{4} \Big[(\lambda + \mu)({\bf a_q}\cdot{\bf q})^2 + (\lambda + \mu)({\bf b_q}\cdot{\bf q})^2 \\
&&+ \mu (a_{\bf q}^2 + b_{\bf q}^2)q^2\Big].
\end{eqnarray}
The partition function of the system can be calculated as $Z = \prod_{\bf q}' Z_{\bf q}$, where
\begin{equation}
\quad Z_{\bf q} = \int \exp\left(-\frac{U_{\bf q}}{k_BT}\right) d {\bf a_q} d{\bf b_q} = \frac{64 \pi^3 k_B^3T^3}{L^9 q^6 \mu^2 (\lambda + 2\mu)}.
\end{equation}
One finds that the mean of $\delta\rho$ is zero, such that
\begin{equation}
\left<\delta\rho\right> = {\sum_{\bf q}}'\frac{1}{Z_{\bf q}}\int \delta\rho_{\bf q} \exp\left(-\frac{U_{\bf q}}{k_BT}\right) d {\bf a_q} d{\bf b_q}  = 0,
\end{equation}
and the variance of $\delta\rho$ reads as:
\begin{eqnarray}
\nonumber
\left<\delta\rho^2\right> &=& {\sum_{\bf q}}'\frac{1}{Z_{\bf q}}\int \delta\rho_{\bf q}^2 \exp\left(-\frac{U_{\bf q}}{k_BT}\right) d{\bf a_q}d{\bf b_q} \\
&&= \frac{2k_BT\rho_0^2}{L^3(\lambda + 2\mu)}{\sum_{\bf q}}' \exp\left(-w^2q^2\right).  \label{eq:drho2}
\end{eqnarray}
To carry out the summation in Eq.~(\ref{eq:drho2}), one applies the identity
\begin{equation}
\sum_{n=-\infty}^{\infty} \exp\left(-\frac{4\pi^2 w^2 n^2}{L^2}\right) = \theta_3\bigl(e^{-4\pi^2 w^2/L^2}\bigr),
\end{equation}
where $\theta_3(x)$ is the third Jacobi theta-function, which permits to calculate $\left<\delta\rho^2\right>$ as
\begin{equation}
\left<\delta\rho^2\right> = \frac{k_BT\rho_0^2}{L^3(\lambda + 2\mu)}\left(\theta_3^3\bigl(e^{-4\pi^2 w^2/L^2}\bigr) - 1\right).
\end{equation}
This equation permits to define the P-wave modulus $M = \lambda + 2\mu$ by using the variance of the local density fluctuations as
\begin{equation}
    M = \frac{k_BT \rho_0^2}{L^3\left<\delta\rho^2\right>}\left(\theta_3^3\bigl(e^{-4\pi^2 w^2/L^2}\bigr) - 1\right).  \label{eq:homM}
\end{equation}
On the spatial scales $w\ll L$ one can use the asymptotics $\theta_3(e^{-x}) \simeq \sqrt{\pi/x}$ for small positive $x$, and obtain the P-wave modulus in the next form:
\begin{equation}
    M = \frac{1}{8\pi^{3/2} }\frac{k_BT \rho_0^2}{w^3\left<\delta\rho^2\right>}.  \label{eq:homM2}
\end{equation}
Equations (\ref{eq:homM}) and (\ref{eq:homM2}) obtained for a homogeneous solid can be used as a definition of the local modulus of an inhomogeneous solid (see Eqs.~(\ref{eq:M}) and (\ref{eq:M2}) in the main text), since it is determined by the local density fluctuations $\left<\delta\rho^2\right>$.

\section{Smooth histogram method}
\label{sec:histogram}

\begin{figure}[t]
    \centering
    \includegraphics[scale=0.9]{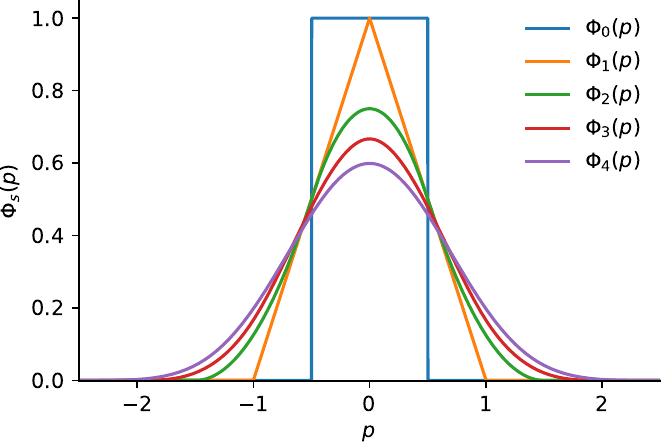}
    \caption{Window functions $\Phi_s(p)$ for different values of $s$.}
    \label{fig:window}
\end{figure}

To obtain the smooth relative density $\xi({\bf r}, t)$, one introduces the three-dimensional histogram
\begin{equation}
h_{ijk} = \sum_n \Phi_s(x_n/d - i)\Phi_s(y_n/d - j)\Phi_s(z_n/d - k), \label{eq:hist}
\end{equation}
where $(x_n, y_n, z_n)$ are coordinates of the $n$th particle in the simulation box (coarse-grained polystyrene particle or nanoparticle atom) and $d=L/N_{\rm bins}$ is the size of a sampling bin in each direction.

Smooth histogram (\ref{eq:hist}) is determined by the window function $\Phi_s(p)$, which was obtained recursively by applying the convolution $\Phi_s(p) = \Phi_{s-1}(p)*\Phi_{0}(p)$ (Fig.~\ref{fig:window}). Here $\Phi_{0}(p)$ is the standard box function, which is equal to 1 for $-1/2<p<1/2$ and zero otherwise. The resulting window function $\Phi_s(p)$ is $s-1$ times differentiable and has the following important properties
\begin{gather}
    \sum_i \Phi_s(p - i) = 1, \quad s \geq 0,\\
    \sum_i (p - i)\Phi_s(p - i) = 0, \quad s \geq 1, \\
    \sum_i (p - i)^2 \Phi_s(p - i) = (s+1)/12, \quad s \geq 2
\end{gather}
for any real value $p$. For $s\gg 1$ the window function $\Phi_s(p)$ is close to a Gaussian with the variance equal to $(s+1)/12$. To obtain the best balance between the accuracy and the calculation time, one can employ $s=4$ and an additional Gaussian smoothing of $h_{ijk}$ with the variance $\sigma^2 = (w/d)^2 - (s+1)/12$ using the fast Fourier transform (FFT). In this case the window function is
\begin{equation}
    \Phi_4(p) = \frac{1}{384}
    \begin{cases}
        (5 + 2 p)^4, & -\frac{5}{2}<p\leq -\frac{3}{2}, \\
        (5 + 2 p)^4-5 (3 + 2 p)^4, & -\frac{3}{2}<p\leq -\frac{1}{2}, \\
        96 p^4-240 p^2+230, & -\frac{1}{2}<p\leq \frac{1}{2}, \\
        (5-2 p)^4-5 (3-2 p)^4, & \frac{1}{2}<p\leq \frac{3}{2}, \\
        (5-2 p)^4, & \frac{3}{2}<p\leq \frac{5}{2}, \\
        0, & \text{elsewhere},
       \end{cases}
\end{equation}
which is smooth and close to a Gaussian with the variance $5/12$. In the present study, we assume $N_{\rm bins} = 100$, which is enough for the studied values of the smoothing parameter $w$.

\section{Distribution of elastic properties}

\begin{figure}[h]
    \centering
    \includegraphics[scale=0.9]{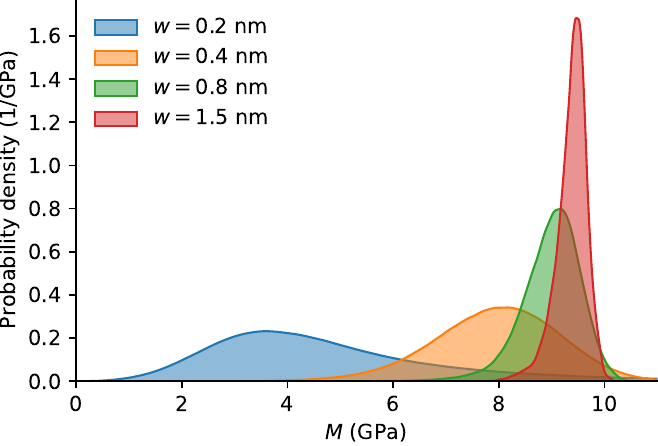}
    \caption{Probability density function for local elastic modulus $M({\bf r})$ in polystyrene for different spatial scales $w=0.2, 0.4, 0.8, 1.5$ nm.} 
    \label{fig:hist}
\end{figure}

Figure~\ref{fig:hist} shows the distribution of the local elastic modulus $M({\bf r})$ in pure polystyrene for different spatial scales $w=0.2, 0.4, 0.8, 1.5$ nm. The last three cases correspond to the local elasticity maps given in Fig.~\ref{fig:map_PS}(a--c).

\section{Elastic properties of the nanoparticle}

\begin{figure}[h]  
    \centering
    \includegraphics[scale=0.9]{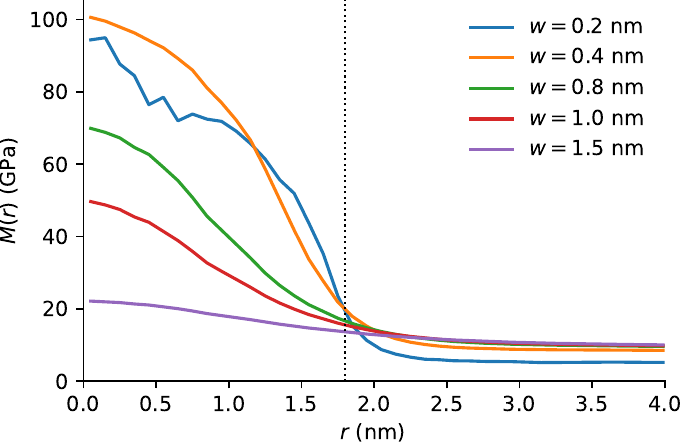}
    \caption{Local elastic modulus $M({\bf r})$ in the nanoparticle and the surrounding polystyrene as a function of the distance to the center of the nanoparticle. The vertical dotted line indicates the boundary of the nanoinclusion.}
    \label{fig:M_NP}
\end{figure}

Figure~\ref{fig:M_NP} shows the local elastic modulus $M({\bf r})$ in the nanoparticle and the surrounding polystyrene. For small values of the spatial scale ($w=0.2$ nm and $w= 0.4$ nm), the obtained values of the P-wave modulus $M$ inside the nanoparticle are in agreement with the typical values for amorphous SiO$_2$ ($M=77$ GPa~\cite{Deschamps-2014}) and quartz ($M=97$ GPa~\cite{Mavko-1995}). The obtained values of $M({\bf r})$ near the boundary of the nanoparticle are suppressed by the vicinity of the relatively soft polystyrene matrix due to the smoothing scale $w$. For $w\geq 0.8$ nm the smoothing suppresses $M({\bf r})$ in the entire volume of the nanoparticle.
 
\section{Structural properties of polystyrene around the nanoparticle}
\label{sec:structure}

\begin{figure}[!t]
    \centering
    \includegraphics[scale=0.9]{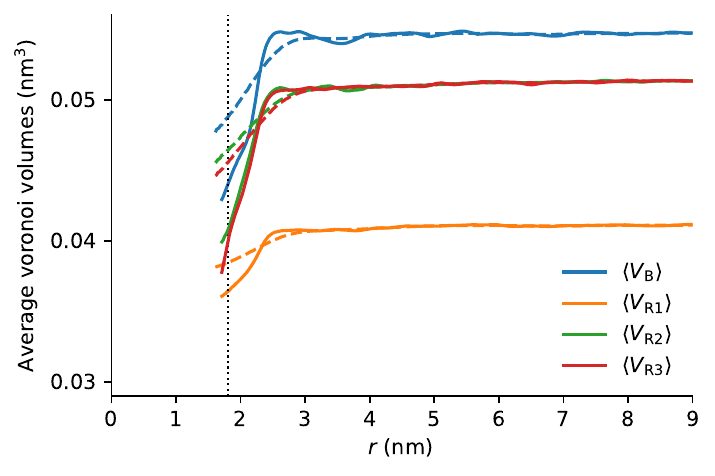}
    \caption{Distribution of the Voronoi volumes as a function of distance from the center of the nanoparticle. Different colors represent different types of the coarse-grain particles: B, R1, R2, R3. Solid and dashed lines show the result for the length scales $w=0.1$ nm and $w=0.4$ nm, respectively. The vertical dotted line indicates the boundary of the nanoparticle. }
    \label{fig:Voronoi}
\end{figure}

\begin{figure}[!htp]
    \centering
    \includegraphics[scale=0.9]{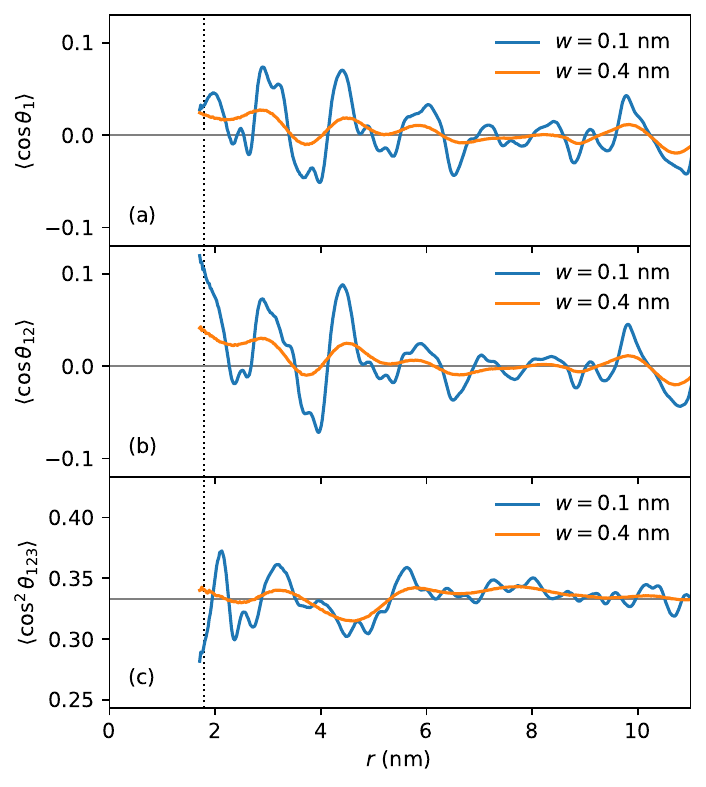}
    \caption{Anisotropy parameters of the orientation of monomers $\langle\cos\theta_1\rangle$, $\langle\cos\theta_{12}\rangle$, and $\langle\cos^2\theta_{123}\rangle$ as a function of the distance to the center of the nanoparticle. The vertical dotted line indicates the boundary of the nanoparticle. Horizontal gray lines indicate the average values for isotropic random orientation: 0, 0, and 1/3, respectively. }
    \label{fig:Angles}
\end{figure}

Here we study additional properties of polystyrene near the nanoparticle to show that the structural properties of the simulated polystyrene do not depend on the distance from the nanoparticle except for a very thin interfacial layer.

In addition to the polystyrene density, we study the spatial distribution of individual Voronoi volumes around coarse-grain particles B, R1, R2, R3. For particles of type B, this spatial distribution has the following form
\begin{equation}
\langle V_{\rm B}\rangle(r) = \frac{\big\langle \sum_{i\in \{{\rm B}\}} V_i \phi(\mathbf{r}_i - \mathbf{r}) \big\rangle}{\big\langle \sum_{i\in \{{\rm B}\}} \phi(\mathbf{r}_i - \mathbf{r})\big\rangle}.
\label{eq:Voronoi}
\end{equation}
Here the averaging is performed over the simulation time $t$ and the coordinate ${\bf r}$ for the given distance to the center of the nanoparticle $r$ (we assume that the origin of the used coordinate frame is at the center of the nanoparticle). The smoothing function $\phi$ in Eq.~(\ref{eq:phi}) defines the spatial scale $w$. Similar definitions $\langle V_{\rm R1}\rangle(r)$, $\langle V_{\rm R2}\rangle(r)$, $\langle V_{\rm R3}\rangle(r)$ for types R1, R2, R3, respectively, can also be adopted from Eq.~(\ref{eq:Voronoi}). The results of the calculated average Voronoi volumes are presented in Fig.~\ref{fig:Voronoi} as a function of the distance to the center of the nanoparticle for $w=0.1$ nm and $w=0.4$ nm. One can observe the homogeneous distribution of Voronoi volumes for every type except a thin interphase layer around the nanoparticle.

The polymer structure also depends on the orientation of the monomers. For each monomer, the vector between the particle B on the polymer's backbone and the central particle R1 in the monomer (see inset in Fig.~\ref{fig:PSMD}(a)) defines as follows:
\begin{equation}
    {\bf r}_{1} = {\bf r}_{\rm R1} - {\bf r}_{\rm B}.
\end{equation}
The vector between the central particle R1 and side particle R2 in the monomer is
\begin{equation}
    {\bf r}_{12} = {\bf r}_{\rm R2} - {\bf r}_{\rm R1},
\end{equation}
and the vector, which is normal to the triangle R1--R2--R3, is
\begin{equation}
    {\bf n}_{123} = ({\bf r}_{\rm R2} - {\bf r}_{\rm R1})\times ({\bf r}_{\rm R3} - {\bf r}_{\rm R1}).
\end{equation}
The orientation of the monomer may be anisotropic due to the influence of the nanoparticle. If this anisotropy exists, its orientation depends on the vector ${\bf r}_{\rm R1}$ between the monomer and the center of the nanoparticle. For each monomer, the cosines of the angles between the vectors ${\bf r}_{1}$, ${\bf r}_{12}$, ${\bf n}_{123}$ and the vector ${\bf r}_{\rm R1}$ had calculated:
\begin{align}
\cos\theta_1 &= \frac{{\bf r}_1\cdot {\bf r}_{\rm R1} }{r_1 r_{\rm R1}},\\
\cos\theta_{12} &= \frac{{\bf r}_{12}\cdot {\bf r}_{\rm R1} }{r_{12} r_{\rm R1}},\\
\cos\theta_{123} &= \frac{{\bf n}_{123}\cdot {\bf r}_{\rm R1} }{n_{123} r_{\rm R1}}.
\end{align}
The corresponding average values represent the anisotropy of the monomer orientation:
\begin{align}
\langle\cos\theta_1\rangle(r) &= \frac{\big\langle \sum_{i\in \{{\rm R1}\}} \cos\theta_1 \phi(\mathbf{r}_i - \mathbf{r}) \big\rangle}{\big\langle \sum_{i\in \{{\rm R1}\}} \phi(\mathbf{r}_i - \mathbf{r})\big\rangle}, \\
\langle\cos\theta_{12}\rangle(r) &= \frac{\big\langle \sum_{i\in \{{\rm R1}\}} \cos\theta_{12} \phi(\mathbf{r}_i - \mathbf{r}) \big\rangle}{\big\langle \sum_{i\in \{{\rm R1}\}} \phi(\mathbf{r}_i - \mathbf{r})\big\rangle}, \\
\langle\cos^2\theta_{123}\rangle(r) &= \frac{\big\langle \sum_{i\in \{{\rm R1}\}} \cos^2\theta_{123} \phi(\mathbf{r}_i - \mathbf{r}) \big\rangle}{\big\langle \sum_{i\in \{{\rm R1}\}} \phi(\mathbf{r}_i - \mathbf{r})\big\rangle}.
\end{align}
Here the averaging is performed over the simulation time $t$ and the coordinate ${\bf r}$ for the given distance from the center of the nanoparticle $r$. In the third case we calculate the average of $\cos^2\theta_{123}$ because the particles R2 and R3 are identical and the average of $\cos\theta_{123}$ is zero. The results are presented in Fig.~\ref{fig:Angles} for two smoothing length scales $w=0.1$ nm and $w=0.4$ nm. One observes that the calculated values are fluctuating near their values for isotropic random orientation: $\langle\cos\theta_1\rangle \approx 0$, $\langle\cos\theta_{12}\rangle \approx 0$, and $\langle\cos^2\theta_{123}\rangle \approx 1/3$.

\section{Local elastic modulus in the random matrix model}


To further study the role of disorder on the local elastic properties of polystyrene with nanoinclusion, we consider the random matrix model, in which the strength of disorder can be easily varied in many orders of magnitude~\cite{Beltukov-2013}. Here we present the essential aspects of this model. 

The random matrix approach under consideration is based on the random nature of the dynamical matrix ${\cal M}$, which describes the motion of particles of a solid near the equilibrium position in harmonic approximation:
\begin{equation}\label{eq:hm}
    \ddot{u}_i^{\alpha}(t) = -\sum_{j\beta}{\cal M}_{ij}^{\alpha\beta} u_j^\beta(t),  
\end{equation}
where ${\bf u}_i(t)=\sqrt{m_i}({\bf r}_i(t) - {\bf r}_i^{(0)})$ describes the scaled deviation of the $i$th particle from its equilibrium position at a time instance $t$. The dynamical matrix is defined by the second derivatives of the total potential energy $U$ at the equilibrium as ${\cal M}_{ij}^{\alpha\beta} = (m_im_j)^{-1/2}\partial^2 U / \partial r_i^\alpha \partial r_j^\beta$. For simplicity, we consider the case of unit masses $m_i=1$ for all particles in the system. Also, we assume that the deviations of particles from their equilibrium position are described by the scalar quantities $u_i$, which are much smaller than a typical distance between particles. In this so-called scalar model, the particles move along one direction only (e.g.\ in the $z$ direction), such that ${\bf r}_i(t) = {\bf r}^{(0)}_i + u_i(t){\bf e}_z$, so the spatial indices $\alpha$ and $\beta$ of the dynamical matrix (\ref{eq:hm}) can be omitted.

In disordered solids, the elements of the dynamical matrix ${\cal M}_{ij}$ have a random nature. However, there are correlations between different elements of the dynamical matrix. Some of these correlations are determined by the particular atomic structure of the solid under consideration. At the same time, there are two fundamental relations between the matrix elements, which should be satisfied for any disordered solid near equilibrium: (i) the system is near the stable equilibrium position and (ii) the potential energy is invariant under the continuous translation of the system~\cite{Conyuh-2021}. In the harmonic approximation given by Eq.~(\ref{eq:hm}), the mechanical stability (i) corresponds to the non-negativity of eigenvalues of ${\cal M}$, which are the squares of eigenfrequencies. The property (ii) leads to the sum rule ${\cal M}_{ii} = -\sum_{j\neq i} {\cal M}_{ji}$. The two properties above can be satisfied if the dynamical matrix is written as~\cite{Beltukov-2013}:
\begin{equation}
    {\cal M} = {\cal A}{\cal A}^T + \mu {\cal M}^{(c)}.
    \label{eq:RMT} 
\end{equation}
The matrix ${\cal A}{\cal A}^T$ describes a disorder in the system. We consider a simple force-constant disorder with particles being at the sites of a simple cubic lattice with unit lattice constant $a_0=1$. The disorder is defined by the random zero-mean Gaussian elements ${\cal A}_{ij}$ with variance $\langle {\cal A}_{ij}^2 \rangle = \Omega^2$ for the neighboring particles with indices $i$ and $j$, where the constant $\Omega$ defines a typical frequency in the system. Other non-diagonal elements of the matrix ${\cal A}$ are zero. The diagonal elements obey the sum rule: ${\cal A}_{ii} = -\sum_{j\neq i} {\cal A}_{ji}$ to satisfy the sum rule (ii) for the dynamical matrix ${\cal M}$. The multiplication ${\cal A}{\cal A}^T$ guarantees the mechanical stability of the system for any random elements ${\cal A}_{ij}$~\cite{Beltukov-2013}. 

Additional matrix ${\cal M}^{(c)}$ in Eq.~(\ref{eq:RMT}) corresponds to the ordered contribution in the dynamical matrix. For the neighboring particles $i$ and $j$, ${\cal M}_{ij}^{(c)}=-\Omega^2$, while the other non-diagonal elements are zero. Due to the sum rule, the diagonal elements are ${\cal M}_{ii}^{(c)} = 6\Omega^2$ for a simple cubic lattice. The heterogeneity of the system is controlled by the dimensionless parameter $\mu \geq 0$ which characterizes the ratio between order and disorder in the system.

The analysis of the local density fluctuations was applied to calculate the local elastic modulus $M({\bf r})$ within the framework of the random matrix model. To simulate the behavior at some temperature $T$, one can solve the Langevin equation
\begin{equation}
    \ddot{u}_i(t) = -\sum_{j\beta}{\cal M}_{ij}u_j(t) - \gamma \dot{u}_i(t) + \eta_i(t),   \label{eq:RMT_MD}
\end{equation}
where $\gamma$ is a damping constant and $\eta_i(t)$ is the noise term defined through the delta-correlated random Gaussian forces acting on each particle in the system:
\begin{equation}
    \langle \eta_i(t)\rangle = 0, \quad \langle \eta_i(t) \eta_j(t') \rangle = 2 \gamma k_\textsc{b} T \delta_{ij}\delta(t - t').
\end{equation}
Here $T$ is the temperature and $k_B$ is the Boltzmann constant. In the employed model, a harmonic approximation was used to describe the motion of particles, hence the elastic moduli do not depend on the temperature. The Langevin equation (\ref{eq:RMT_MD}) have been solved using the stochastic Verlet method~\cite{Gronbech-2020} with an integration timestep $\Delta t= 0.02 \Omega^{-1}$ and the damping coefficient $\gamma=0.1 \Omega$.

The local density fluctuations were then calculated as
\begin{equation}
    \xi({\bf r}, t) = \sum_i u_i \phi'({\bf r} - {\bf r}_i^{(0)}),
\end{equation}
where $\phi'({\bf r}) = \partial \phi ({\bf r})/\partial z$ is the derivative of the smoothing function $\phi$ since $u_i$ represents a small deviation of the $i$th particle from its equilibrium position ${\bf r}_i^{(0)}$ in the $z$ direction. One can easily generalize the model to a three-dimensional vector case, where all the three principal directions would be involved. However, the considered scalar model is sufficient to emphasize the role of disorder on the local elastic properties. 

\begin{figure}[!htp]
    \centering
    \includegraphics[scale=0.9]{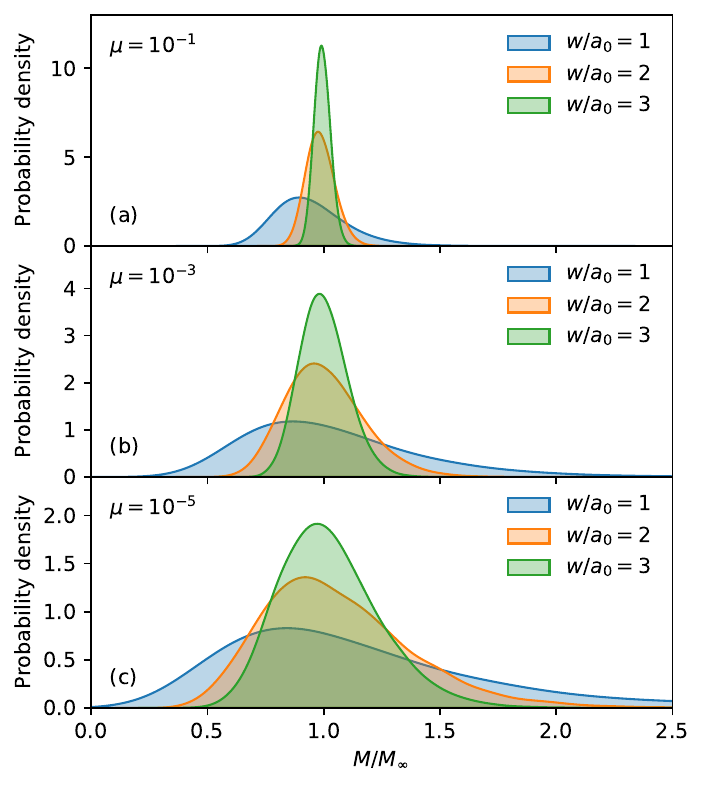}
    \caption{Probability density of the local elastic modulus in the random matrix model for three different values of the parameter $\mu=10^{-1}, 10^{-3}, 10^{-5}$ and the three different spatial scales $w/a_0=1,2,3$. The elastic moduli are normalized by $M_\infty$, which is the mean elastic modulus for $w\to\infty$.}
    \label{fig:hist_RMT}
\end{figure}

The local elastic modulus $M({\bf r})$ was calculated using Eq.~(\ref{eq:M}) for different spatial scales $w$ and different values of the parameter $\mu$. With the increase of the spatial scale $w$, the local elastic modulus $M({\bf r})$ tends to its macroscopic value $M_\infty$ ($M_\infty \propto \sqrt{\mu}$ for $\mu\ll 1$, see~\cite{Beltukov-2013}). For a finite spatial scale $w$, there is a broadening of the probability density of $M({\bf r})$, as shown in Fig.~\ref{fig:hist_RMT}. One can see that the relative broadening of $M({\bf r})/M_{\infty}$ increases with decreasing $\mu$ for a given spatial scale $w$,  emphasizing the fact that disorder of the elastic properties grows as the value of the parameter $\mu$ decreases.

Usually nanoinclusions have a more rigid and ordered structure than the surrounding material. Therefore, for a simulation of a nanoinclusion in the random matrix model, a spherical area $|{\bf r}_i^{(0)}| < R_\textsc{np}$ was chosen with the parameter $\mu = 1$ set for that region. The parameter $\mu\ll 1$ of the surrounding medium can be varied. In the main paper in Fig.~\ref{fig:dE} we show that the presence of a nanoinclusion increases the elastic modulus by $M(r)/M_0 - 1 \sim \exp(-r/\lambda)$ with $\lambda\sim \mu^{-1/4}$. The scaling of $\lambda$ coincides with the known scaling of the Ioffe-Regel length in the random matrix model~\cite{Beltukov-2013}. 

\begin{figure}[t]
    \centering
    \includegraphics[scale=0.9]{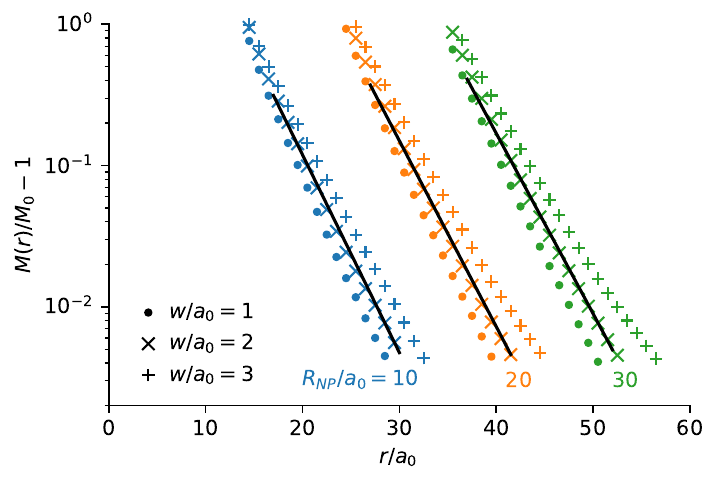}
    \caption{Relative increase of the elastic modulus $M$ near the nanoinclusion computed within the random matrix model for $\mu=10^{-4}$ for different values of the spatial scale: $w=1a_0$ (dots), $w=2a_0$ (diagonal crosses), $w = 3a_0$ (vertical crosses). Different colors correspond to different sizes of nanoinclusion $R_{\textsc{np}}=10a_0$, $20a_0$, $30a_0$. Solid lines show the exponential law $\sim\exp(-r/\lambda)$ for the corresponding values of $R_{\textsc{np}}$. } 
    \label{fig:dE_RNP}
\end{figure}

To emphasize the invariance of the length scale $\lambda$ on the nanoinclusion size, we present the relative increase of the elastic modulus $M$ for different sizes of nanoinclusion. For $\mu=10^{-4}$, the following values of the length scale $\lambda$ were obtained: $\lambda = 3.1a_0, 3.3a_0, 3.4a_0$  for $R_\textsc{np} = 10a_0$, $20a_0$, $30a_0$, respectively (Fig.~\ref{fig:dE_RNP}). Therefore, the thickness $\lambda$ of the induced elastic shell is almost indifferent to the size of the nanoinclusion.

\end{document}